\documentclass{article}

\usepackage{PRIMEarxiv}
\usepackage[utf8]{inputenc} % allow utf-8 input
\usepackage[T1]{fontenc}    % use 8-bit T1 fonts
\usepackage{hyperref}       % hyperlinks
\usepackage{url}            % simple URL typesetting
\usepackage{booktabs}       % professional-quality tables
\usepackage{amsfonts}       % blackboard math symbols
\usepackage{nicefrac}       % compact symbols for 1/2, etc.
\usepackage{microtype}      % microtypography
\usepackage{lipsum}
\usepackage{fancyhdr}       % header
\usepackage{graphicx}       % graphics
\graphicspath{{media/}}     % organize your images and other figures under media/ folder

\usepackage[table]{xcolor}
\usepackage{xspace}
\usepackage{multirow}
\usepackage{comment}
\usepackage{enumitem}
\usepackage[xindy,acronym]{glossaries}
\glsdisablehyper

\newcommand{\lsrr}{\textit{LSRR}\xspace}
\newcommand{\ldr}{\textit{LDR}\xspace}
\newcommand{\er}{\textit{TIAGo Pro}\xspace}

\newcommand{\fasterrcnn}{\textit{Faster R-CNN}\xspace}
\newcommand{\fasterrcnnv}{\textit{Faster R-CNN\_v2}\xspace}
\newcommand{\retinanet}{\textit{RetinaNet}\xspace}
\newcommand{\retinanetv}{\textit{RetinaNet\_v2}\xspace}
\newcommand{\ssd}{\textit{SSD300}\xspace}
\newcommand{\ssdlite}{\textit{SSDLite}\xspace}
\newcommand{\stdm}{\textit{STDM}\xspace}
\newcommand{\scdm}{\textit{SCDM}\xspace}
\newcommand{\hfdm}{\textit{HFDM}\xspace}

\newcommand{\uamters}{\textit{UAMTERS}\xspace}
\newcommand{\match}{\emph{Match}\xspace}
\newcommand{\miss}{\emph{Miss}\xspace}
\newcommand{\ghost}{\emph{Ghost}\xspace}
\newcommand{\mcd}{\textit{MCD}\xspace}
\newcommand{\mcb}{\textit{MCB}\xspace}

\newcommand{\origimg}{\textit{origImg}\xspace}
\newcommand{\dalleimg}{\textit{daImg}\xspace}
\newcommand{\sdimg}{\textit{sdImg}\xspace}

\newcommand{\normalimg}{\textit{norImg}\xspace}
\newcommand{\diffexpimg}{\textit{expImg}\xspace}

\newcommand{\lowimg}{\textit{lowImg}\xspace}
\newcommand{\mediumimg}{\textit{medImg}\xspace}
\newcommand{\highimg}{\textit{highImg}\xspace}

\newcommand{\imgms}{\textit{Img-MS}\xspace}
\newcommand{\objms}{\textit{Obj-MS}\xspace}
\newcommand{\ioums}{\textit{IoU-MS}\xspace}
\newcommand{\uams}{\textit{UA-MS}\xspace}

\newcommand{\imgmsmiss}{$MS_{img}^{miss}$\xspace}
\newcommand{\imgmsghost}{$MS_{img}^{ghost}$\xspace}
\newcommand{\imgmsmg}{$MS_{img}^{mg}$\xspace}

\newcommand{\objmsmiss}{$MS_{obj}^{miss}$\xspace}
\newcommand{\objmsghost}{$MS_{obj}^{ghost}$\xspace}
\newcommand{\objmsmg}{$MS_{obj}^{mg}$\xspace}

\newcommand{\msiou}{$MS_{iou}$\xspace}

\newcommand{\uamsmatchvr}{$MS_{vr}^{match}$\xspace}
\newcommand{\uamsmatchie}{$MS_{ie}^{match}$\xspace}
\newcommand{\uamsmatchmi}{$MS_{mi}^{match}$\xspace}
\newcommand{\uamsmatchva}{$MS_{va}^{match}$\xspace}
\newcommand{\uamsmatchps}{$MS_{ps}^{match}$\xspace}

\newcommand{\uamsmissvr}{$MS_{vr}^{miss}$\xspace}
\newcommand{\uamsmissie}{$MS_{ie}^{miss}$\xspace}
\newcommand{\uamsmissmi}{$MS_{mi}^{miss}$\xspace}
\newcommand{\uamsmissva}{$MS_{va}^{miss}$\xspace}
\newcommand{\uamsmissps}{$MS_{ps}^{miss}$\xspace}

\newcommand{\uamsghostvr}{$MS_{vr}^{ghost}$\xspace}
\newcommand{\uamsghostie}{$MS_{ie}^{ghost}$\xspace}
\newcommand{\uamsghostmi}{$MS_{mi}^{ghost}$\xspace}
\newcommand{\uamsghostva}{$MS_{va}^{ghost}$\xspace}
\newcommand{\uamsghostps}{$MS_{ps}^{ghost}$\xspace}
\newcommand{\EUProject}{{\tt RoboSAPIENS}\xspace}

\definecolor{corrLow}{RGB}{255,237,213}
\definecolor{corrModerate}{RGB}{255,204,153}
\definecolor{corrHigh}{RGB}{255,153,51}
\definecolor{corrVeryHigh}{RGB}{230,92,0}

\newtheorem{definition}{Definition}[subsection] % resets numbering per section

\newacronym{ms}{MS}{mutation score}
\newacronym{odm}{ODM}{object detection model}
\newacronym{ts}{TS}{test suite}
\newacronym{ci}{CI}{confidence interval}
\newacronym{dl}{DL}{deep learning}
\newacronym{uq}{UQ}{uncertainty quantification}

\newacronym{vr}{VR}{Variation Ratio}
\newacronym{se}{SE}{Shannon Entropy}
\newacronym{mi}{MI}{Mutual Information}
\newacronym{tv}{TV}{Total Variance}
\newacronym{ps}{PS}{Predictive Surface}

\title{UAMTERS: Uncertainty-Aware Mutation Analysis for DL-enabled Robotic Software
}

\author{
  Chengjie Lu \\
  Simula Research Laboratory and University of Oslo \\
  Oslo \\
  Norway \\
  \texttt{chengjielu@simula.no} \\
  \And
  Jiahui Wu \\
  Simula Research Laboratory and University of Oslo \\
  Oslo \\
  Norway \\
  \texttt{jiahui@simula.no} \\
  \And
  Shaukat Ali \\
  Simula Research Laboratory \\
  Oslo \\
  Norway \\
  \texttt{shaukat@simula.no} \\
  \And
  Malaika Din Hashmi \\
  Danish Technological Institute \\
  Odense \\
  Denmark \\
  \texttt{mdih@teknologisk.dk} \\
  \And
  Sebastian Mathias Thomle Mason \\
  Danish Technological Institute \\
  Odense \\
  Denmark \\
  \texttt{sem@teknologisk.dk} \\
  \And
  Francois Picard \\
  Danish Technological Institute \\
  Odense \\
  Denmark \\
  \texttt{fpi@teknologisk.dk} \\
  \And
  Mikkel Labori Olsen \\
  Danish Technological Institute \\
  Odense \\
  Denmark \\
  \texttt{miol@teknologisk.dk} \\
  \And
  Thomas Peyrucain \\
  PAL Robotics \\
  Barcelona \\
  Spain \\
  \texttt{thomas.peyrucain@pal-robotics.com} \\
}

\begin{document}
\maketitle

\begin{abstract}
Self-adaptive robots adjust their behaviors in response to unpredictable environmental changes. These robots often incorporate deep learning (DL) components into their software to support functionality such as perception, decision-making, and control, enhancing autonomy and self-adaptability. 
However, the inherent uncertainty of DL-enabled software makes it challenging to ensure its dependability in dynamic environments. Consequently, test generation techniques have been developed to test robot software, and classical mutation analysis injects faults into the software to assess the test suite's effectiveness in detecting the resulting failures. However, there is a lack of mutation analysis techniques to assess the effectiveness under the uncertainty inherent to DL-enabled software. To this end, we propose \uamters, an uncertainty-aware mutation analysis framework that introduces uncertainty-aware mutation operators to explicitly inject stochastic uncertainty into DL-enabled robotic software, simulating uncertainty in its behavior. We further propose mutation score metrics to quantify a test suite's ability to detect failures under varying levels of uncertainty. We evaluate \uamters across three robotic case studies, demonstrating that \uamters more effectively distinguishes test suite quality and captures uncertainty-induced failures in DL-enabled software.
\end{abstract}

% keywords can be removed

\section{Introduction}
Self-adaptive robots are designed to handle unexpected situations and make decisions in response to unpredictable environmental changes. Such systems have shown great potential in safety-critical domains, including autonomous driving~\cite{zimmermann2020adaptive}, industrial automation~\cite{10.1145/3589227}, and maritime~\cite{wang2025physics}, where operating conditions are highly dynamic and unpredictable at runtime. In recent years, the use of \gls{dl} components in self-adaptive robotic software has increased, as they enable advanced perception, decision-making, and control, thereby enhancing autonomy and adaptability. Meanwhile, integrating \gls{dl} into robotic software increases complexity, making it more challenging to ensure dependability in unpredictable environments. Moreover, \gls{dl} models inherently exhibit various forms of uncertainty arising from factors such as limited training data and stochastic inference mechanisms (e.g., prediction-time dropout)~\cite{gal2016uncertainty}, further complicating the validation of robot behavior at runtime.

To ensure the dependability of robotic software, various testing techniques have been proposed~\cite{karoly2020deep}. However, there is a lack of mutation analysis approaches for assessing the quality of generated test suites, particularly from the perspective of their ability to deal with the inherent uncertainty of \gls{dl} components in robotic software. Such mutation analysis techniques help evaluate the ability of test suites to reliably detect safety and performance violations under realistic and uncertain conditions.
Various mutation analysis approaches have been proposed for \gls{dl}-enabled robotic software. For example, DeepMutation~\cite{ma2018deepmutation} and DeepMutation++~\cite{hu2019deepmutation++} inject faults into \gls{dl} models using novel mutation operators and can be applied across a wide range of \gls{dl} architectures. Furthermore, DeepCrime~\cite{humbatova2021deepcrime} integrates 35 mutation operators derived from the analysis of real \gls{dl} faults, enabling the simulation of real faults in \gls{dl} models. In addition, Jahangirova and Tonella~\cite{jahangirova2020empirical} conduct an empirical evaluation to identify effective mutation operators and propose a statistical definition of mutation killing, further supporting rigorous mutation testing of \gls{dl}-enabled systems.

While the above works advance mutation analysis for \gls{dl}-enabled software, they do not explicitly consider the uncertainty inherent in \gls{dl} models. Uncertainty arising from limited training data, dropout, and stochastic inference mechanisms can significantly affect robot behavior, particularly in complex and unpredictable operating environments. 
As a closely related work, DeepCrime~\cite{humbatova2021deepcrime} introduces a mutation operator that modifies dropout layers; it treats dropout as a training-time regularization technique rather than as a mechanism for characterizing predictive uncertainty during inference. 
Therefore, there is a need for mutation analysis techniques to assess the ability of test suites to detect failures arising from uncertainty in \gls{dl} components within robotic software.

To address this limitation, we propose \uamters, an uncertainty-aware mutation analysis framework, for \gls{dl}-enabled software in robots. \uamters proposes novel operators that inject uncertainty into \gls{dl} models, enabling systematic simulation of uncertain model behaviors to assess the test suite's ability to reveal failures caused by these behaviors. We quantify this with novel uncertainty-aware mutation score metrics. By explicitly accounting for uncertainty, \uamters enables a more informative evaluation of test suite quality in revealing failures caused by uncertainty. We evaluate \uamters using three industrial-level robotic case studies, comprising seven embedded \gls{dl} models and eight test suites. The results show that our novel mutation score metrics are more effective in distinguishing test suite quality than traditional mutation scores. Moreover, the proposed mutation score metrics capture model degradation by tracking changes in uncertainty.

In summary, our contributions are: 
1) \uamters, an uncertainty-aware mutation analysis framework for \gls{dl}-enabled robotic software integrating novel uncertainty-aware mutation operators to explicitly inject stochastic uncertainty into robot behaviors;
2) Novel uncertainty-aware mutation score metrics that quantify a test suite's ability to reveal model degradation and behavioral changes under varying levels of uncertainty;
3) Extensive empirical evaluation across industrial-level robotic case studies, demonstrating that the proposed mutation score metrics more effectively distinguish test suite quality and capture uncertainty-induced model performance degradation.

\section{Context and Background}

\subsection{Context: RoboSAPIENS European Project}

\begin{figure}[ht]
    \centering
    \includegraphics[width=0.86\linewidth]{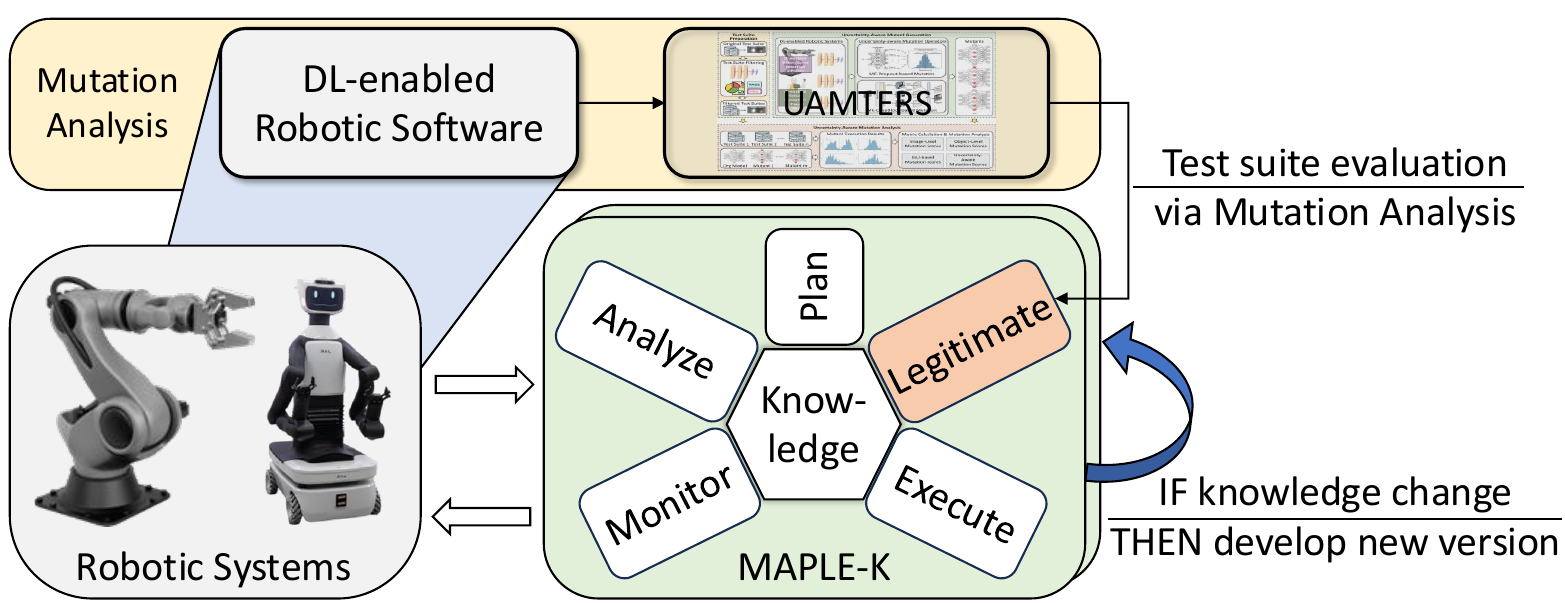}
    \caption{\uamters in the Context of the MAPLE-K Loop.}
    \label{fig:maple-k}
\end{figure}

\EUProject~\cite{larsen2024robotic} is a large European initiative focusing on building open-ended, safe, and trustworthy self-adaptative robotic systems. It extends the well-established self-adaptation framework, the MAPE-K loop~\cite{arcaini2015modeling}, to ensure safe and reliable robot behaviors during operation time. A typical MAPE-K loop consists of five components: the Monitor component continuously observes the system's state, while the Analyze component processes this state information to identify potential needs for adaptation. When adaptation is required, the Plan component generates an appropriate strategy, which is then executed by the Execute component. During this process, all components contribute to the Knowledge component, which stores relevant system information and historical data. The extended framework, referred to as the MAPLE-K loop (see Fig.~\ref{fig:maple-k}), introduces an additional component called Legitimate, which validates the safety and performance of the robot's planned actions before execution. 
A key consideration of Legitimate is the design of test suites that can effectively assess whether planned actions satisfy safety and performance requirements under uncertain and dynamic conditions. Although various test generation techniques have been proposed to evaluate the self-adaptive capabilities of robotic systems~\cite{10.1145/3712256.3726399}, existing approaches largely focus on generating test cases rather than assessing the quality of the resulting test suites. 
In this context, mutation analysis offers a systematic approach for evaluating test suite quality in self-adaptive robotic systems. By introducing controlled mutants into the system, it simulates realistic perturbations in robot behavior. This allows for the evaluation of how effectively a test suite can detect potential failures, supporting the Legitimate component in the MAPLE-K loop.

\subsection{Uncertainty in Deep Learning}\label{subsec:uncer_dl}
\Gls{dl} has shown significant success in fields such as computer vision~\cite{voulodimos2018deep}, natural language processing~\cite{9075398}, and robotics~\cite{soori2023artificial}. However, \gls{dl} still faces major challenges in managing uncertainty, which can limit its reliability in real-world applications~\cite{gal2016uncertainty,gawlikowski2023survey}. Uncertainty in \gls{dl} models is typically categorized into aleatoric uncertainty, which arises from inherent noise or randomness in the data (e.g., sensor errors or poor data quality), and epistemic uncertainty, which is caused by limited knowledge of optimal model parameters and commonly appears when the model encounters previously unseen data. Recent literature~\cite{10.1145/3377811.3380368,9438598} identifies uncertainty in DL as an important cause of failures in \gls{dl} systems, leading to unreliable behavior in safety-critical applications. 
\Gls{uq} is essential for assessing the reliability and trustworthiness of DL models in real-world applications. To this end, various \gls{uq} approaches have been proposed, including Bayesian Neural Networks (BNNs)~\cite{tran2019bayesian}, MC-Dropout~\cite{gal2016dropout}, MC-DropBlock~\cite{yelleni2024monte}, and Deep Ensembles~\cite{lakshminarayanan2017simple}. BNNs provide a probabilistic framework to model uncertainty by applying Bayesian inference to DL models~\cite{tran2019bayesian}; however, their practical application is often limited by the high computational cost and the complexity of posterior inference. As a result, several Bayesian approximation methods have been proposed to enable more efficient and scalable \gls{uq}. Gal and Ghahramani~\cite{gal2016dropout} introduce MC-Dropout, which approximates Bayesian inference by retaining dropout layers during inference and performing multiple stochastic forward passes. 
Building on this idea, MC-DropBlock~\cite{yelleni2024monte} extends MC-Dropout to convolutional neural networks (CNNs) by randomly dropping contiguous regions in feature maps, thereby providing a more structured and effective \gls{uq} for CNNs.

Beyond \gls{uq} methods, designing effective \gls{uq} metrics is important for capturing uncertainties in \gls{dl} models. Focusing on DL-enabled object detection software, Lu et al.~\cite{lu2025assessing} implement two categories of \gls{uq} metrics: \gls{uq} metrics for label classification and \gls{uq} metrics for bounding box regression. For classification tasks, three common metrics are used~\cite{gal2016uncertainty}: \gls{vr}~\cite{freeman1965elementary}, \gls{se}~\cite{shannon1948mathematical}, and \gls{mi}~\cite{shannon1948mathematical}. \gls{vr} measures dispersion by calculating the proportion of predictions that differ from the most frequent class, \gls{se} quantifies the average information in the predictive distribution, and \gls{mi} captures epistemic uncertainty by estimating the discrepancy between the predictive distribution and the posterior over model parameters. For bounding box regression, uncertainty is estimated using \gls{tv}~\cite{feng2018towards} and \gls{ps}~\cite{catak2021prediction}. \gls{tv} sums the variances of the box coordinates from the covariance matrix of the predicted boxes, while \gls{ps} measures spatial uncertainty by computing the area of the convex hull enclosing the predicted bounding box corners.

In this work, we design two uncertainty-aware mutation operators based on the MC-Dropout and MC-DropBlock methods to explicitly inject stochastic uncertainty into \gls{dl}-enabled software. To evaluate test suite effectiveness, we further propose novel uncertainty-aware mutation scores, calculated based on the \gls{uq} metrics employed by Lu et al.~\cite{lu2025assessing}, which quantify the ability of test suites to reveal model degradation under different levels of uncertainty.

\section{Methodology}

\subsection{Overview}
\begin{figure}[ht]
    \centering
    \includegraphics[width=0.98\linewidth]{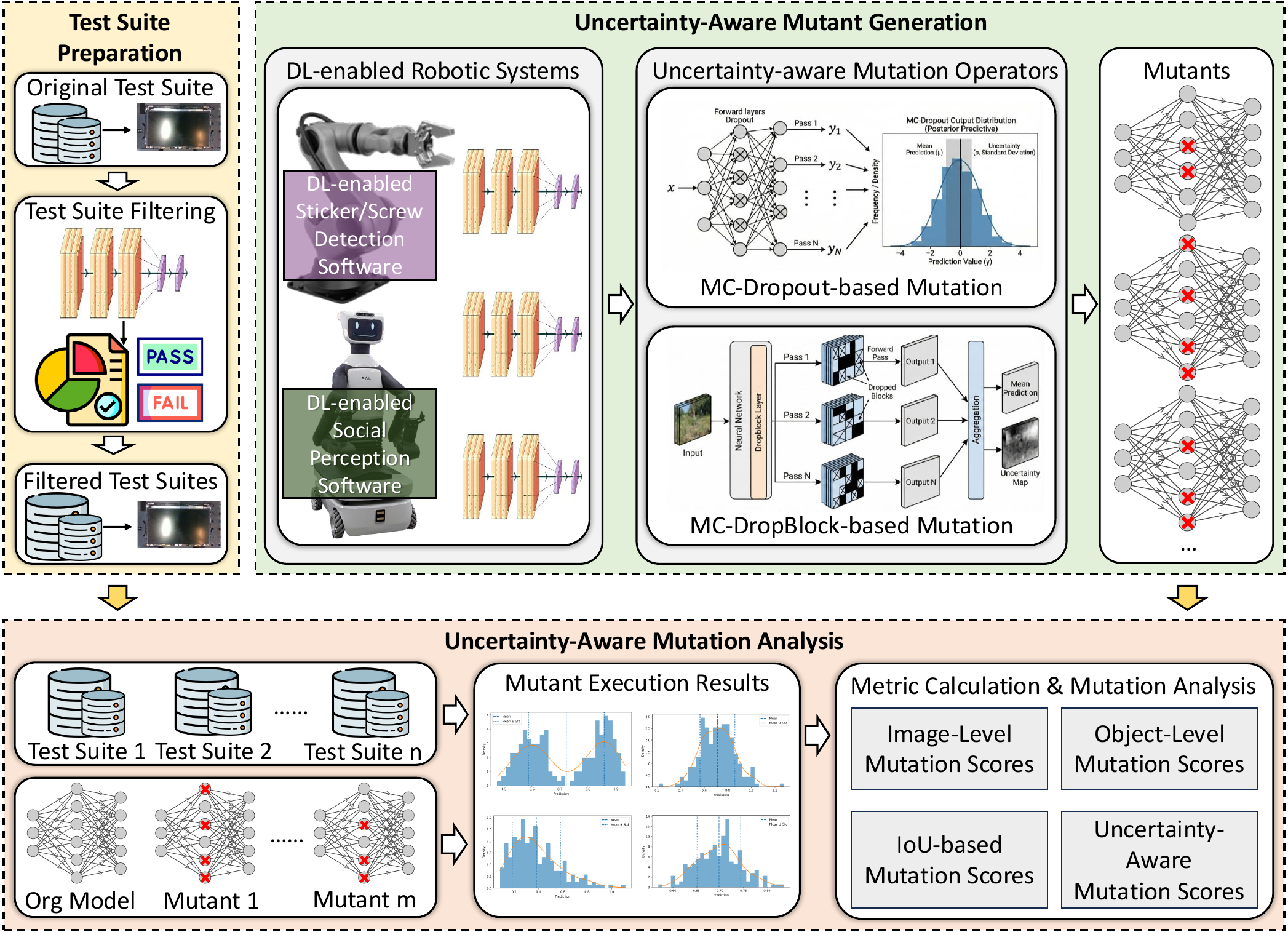}
    \caption{Overview of \uamters.}
    \label{fig:overview}
\end{figure}

Fig.~\ref{fig:overview} presents the workflow of \uamters, which adapts mutation analysis to \gls{dl}-enabled robotic software. The workflow starts with \textit{Test Suite Preparation} to filter out irrelevant test cases. Specifically, given an original test suite \textit{TS}, \uamters executes the original \gls{dl} model \textit{O} on \textit{TS} and keeps the set of passed test cases $TS' \in TS$ for mutation analysis.
After obtaining the \textit{Filtered Test Suites}, the \textit{Uncertainty-Aware Muant Generation} begins, in which the \gls{dl}-enabled robotic software is systematically mutated. In this work, a \textit{Mutant} is defined as a variant of the original \gls{dl} model obtained by applying an uncertainty-aware mutation operator, which injects stochastic uncertainty into the model at inference time.
In detail, we generate mutants of the \gls{dl} components using two mutation operators: MC-Dropout-based mutation (\mcd) and MC-DropBlock-based mutation (\mcb), which are introduced in Section~\ref{subsec:mutation_op}. Finally, in the \textit{Uncertainty-Aware Mutation Analysis} phase, we execute the original models and all their mutants on the \textit{Filtered Test Suites} to evaluate the quality of the test suites based on the execution results. 

\subsection{Uncertainty-Aware Mutation Operators}\label{subsec:mutation_op}
\gls{dl} models may exhibit incorrect behavior due to various factors, among which uncertainty inherent in the model is an important cause, often leading to unstable or incorrect predictions~\cite{d2022underspecification,zhang2020towards}. Motivated by this observation, we design two uncertainty-aware mutation operators to emulate failure scenarios caused by predictive uncertainty in \gls{dl}. These operators generate mutants of the \gls{dl} model by introducing uncertainty into the model, leading to deviations from correct predictive behavior, especially when the model is exposed to unseen input distributions.

\subsubsection{MC-Dropout-based Mutation (\mcd) Operator}
\mcd is a mutation operator that leverages the MC-Dropout method~\cite{gal2016dropout} to inject uncertainty into \gls{dl} models. Since MC-Dropout effectively models uncertainty in \gls{dl} as a Bayesian approximation, we employ it to design the \mcd operator.
Specifically, \mcd works as follows: given a \gls{dl} model $O$, \mcd mutates the model by enabling dropout layers during inference time, resulting in a mutant:
\begin{equation}
\mcd(O, \mu) = O_{\text{mcd}, \mu},
\end{equation}
where $O_{\text{mcd}, \mu}$ denotes the mutated network with inference-time dropout applied, and $\mu$ is the mutation ratio controlling the strength of the mutation. In \mcd, $\mu$ specifies the probability that a neuron is dropped during inference; a higher value of $\mu$ results in more neurons being dropped, thereby injecting greater uncertainty in the model and increasing the chance of uncertainty-induced failures. By adjusting this parameter, \mcd can systematically control the amount of uncertainty injected into the model, enabling the study of model behavior under varying levels of uncertainty.

\subsubsection{MC-DropBlock-based Mutation (\mcb) Operator}
\mcb is a mutation operator designed based on the MC-DropBlock method~\cite{yelleni2024monte}, which was originally proposed to model uncertainty in object detection models. MC-DropBlock is specifically tailored for object detection models with convolutional layers, such as YOLO~\cite{redmon2016you} and convolutional transformers~\cite {wu2021cvt}. Given that our subject systems are object detection models, we employ \mcb to introduce uncertainty into the convolutional layers at inference time, thereby producing mutant models that may exhibit failures induced by predictive uncertainty. Specifically, given a \gls{dl} model $O$, its mutant generated by \mcb is:
\begin{equation}
\mcb(O, \boldsymbol{\mu}) = O_{\text{mcb}, \boldsymbol{\mu}},
\end{equation}
where $\boldsymbol{\mu} = [\mu_d, \mu_b]$ is a vector of mutation ratios: $\mu_d$ is the dropout rate controlling the probability of dropping individual neurons, and $\mu_b$ is the block size controlling the size of contiguous regions of neurons to be dropped. 
Large $\mu_d$ results in more individual neurons being dropped, while larger $\mu_b$ results in bigger contiguous blocks of neurons being dropped in the feature maps. Higher values of these mutation ratios increase the level of uncertainty injected into the model.
By adjusting these two parameters, \mcb can systematically control both the amount and spatial extent of uncertainty injected into the model, enabling the study of model behavior under different levels of uncertainty.

\subsection{Mutation Scores}
Mutation score quantifies the quality of a test suite by measuring how many generated mutants are killed by the test suite. A mutant is considered \textit{killed} if at least one test case in the test suite produces an output that differs from the original model.
For simple DL problems, such as classification and regression, the output is a single vector. When it comes to object detection models, the outputs become a set of bounding boxes, labels, and scores, making the definition of a mutant's behavior change more complex. To address this challenge, we propose a set of novel mutation scores specifically tailored to object detection models. In particular, Section~\ref{subsubsec:objlevel_ms} introduces mutation score metrics defined at the object level, while Section~\ref{subsubsec:uncer_ms} presents novel uncertainty-aware mutation scores. In addition, Section~\ref{subsubsec:imglevel_ms} defines the traditional mutation scores at the image level, which serve as a baseline for comparison with our proposed mutation scores.
Before introducing these metrics, we first formalize the basic concepts required for mutation score calculation. 

\begin{definition}[Output of an Object Dection Model]
Given an object detection model $ODM$ and a test input $img$, the output of $ODM$ is a set of detected objects:
\begin{equation}
    O = ODM(img) = \{d_1, d_2, ..., d_n\},
\end{equation}
where each detection $d_i = (b_i, c_i, s_i, u_i)$ consists of a bounding box coordinates $(x_1, x_2, y_1, y_2)$, a predicted class label $c_i$, and a class probability vector $p_i \in [0, 1]^K$, where each element represents the predicted probablities over the \textit{K} object classes.
\end{definition}

\begin{definition}[\match]
Given a test input $img$ to an original $ODM$ and its mutant $ODM'$, let their outputs be $O$ and $O'$. 
A pair $(d_i, d'_j) \in O \times O'$ is considered a \match if their bounding boxes sufficiently overlap and their labels are the same:
\begin{equation}
    \mathrm{IoU}(b_i, b'_j) > \theta_{\mathrm{IoU}} \land c_i = c'_j,
\end{equation}
where $b_i$ and $b'_j$ denote the bounding boxes associated with detections $d_i$ and $d'_j$, respectively, 
$c_i$ and $c'_j$ denote their corresponding predicted class labels, and $\theta_{\mathrm{IoU}}$ is a predefined threshold. We set $\theta_{\mathrm{IoU}} = 0.5$, following common practice in object detection tasks.
Then, the set of all valid matches is denoted as:
\begin{equation}
    \mathcal{S}_{match} = \{ (d_i, d'_j) \in O \times O' \mid (d_i, d'_j) \text{ satisfies (2)} \}.
\end{equation}
\end{definition}

\begin{definition}[\miss]
A detection $d_i$ in the original output set $O$ is defined as a \miss if it fails to find a valid counterpart in the mutant set $O'$ within the set of matches $\mathcal{M}$. Formally, the set of Misses is defined as:
\begin{equation}
     \mathcal{S}_{miss} = \{ d_i \in O \mid \forall d'_j \in O', (d_i, d'_j) \notin  \mathcal{S}_{match} \}.
\end{equation}
These represent objects correctly identified by the original model but missed by the mutant.
\end{definition}

\begin{definition}[\ghost]
A detection $d'_j$ in the mutant output set $O'$ is defined as a \ghost if it is not paired with any detection in the original set $O$ within the set of matches $\mathcal{M}$. Formally, the set of Ghosts is defined as:
\begin{equation}
     \mathcal{S}_{ghost} = \{ d'_j \in O' \mid \forall d_i \in O, (d_i, d'_j) \notin  \mathcal{S}_{match} \}.
\end{equation}
These represent false positives introduced by the mutation that were not present in the original model's output.
\end{definition}

\subsubsection{Image-level Mutation Score}\label{subsubsec:imglevel_ms}
Image-level mutation scores measure test suite quality by capturing behavioral deviations between the original object detection model and its mutant models at the image (i.e., test case) level. In \gls{dl} mutation analysis,
% DL mutation testing, 
a mutant is considered to be killed if its output differs from that of the original model. Further, to mitigate the effects of randomness inherent in stochastic model behaviors, Jahangirova and Tonella~\cite{jahangirova2020empirical} introduced a statistical notion of mutation killing, in which a mutant is regarded as killed only when the output difference between the original model and the mutant is statistically significant. 
Therefore, we execute the original model and its mutant $n$ times and define a mutant as killed under a test suite $TS$ if the behavioral difference observed on at least one test case in $TS$ is statistically significant across repeated executions. The behavioral difference is characterized in terms of the presence of \miss objects, \ghost objects, and their combination, respectively.
\begin{equation}
\textit{isKilled}(m, \mathcal{S}) =
\begin{cases}
\text{True}, &
\text{if } \exists\, \textit{img} \in TS \text{ such that }
\{\mathbb{I}(|\mathcal{S}^{img,i}| > 0)\}_{i=1}^{n}
\text{ is statistically significant}, \\
\text{False}, & \text{otherwise}.
\end{cases}
\end{equation}
where $\mathbb{I}(\cdot)$ is an indicator function that evaluates to 1 if the set is non-empty and 0 otherwise. We employ the one-sample binomial test to determine whether the observed frequency of non-empty sets across $n$ executions is statistically significantly higher than expected by chance. 
Accordingly, the image-level mutation scores are defined as:
\begin{equation}
    Img\text{-}MS = \frac{|\{m \in M \mid \textit{isKilled}(m, \mathcal{S}) = \text{True}\}|}{|M|},
\end{equation}
where \(M\) is the set of all mutants, \(\mathcal{S} \in \{\mathcal{S}_{miss}, \mathcal{S}_{ghost}, \mathcal{S}_{miss} \cup \mathcal{S}_{ghost}\}\), and \(\textit{isKilled}(m, \mathcal{S})\) indicates whether mutant \(m\) is killed for the set \(\mathcal{S}\) according to the statistical test defined above. Finally, we obtain three image-level \glspl{ms}: \imgmsmiss, \imgmsghost, and \imgmsmg.

\subsubsection{Object-level Mutation Score}\label{subsubsec:objlevel_ms}
Object-level mutation scores evaluate test suite quality by capturing behavioral deviations between the original model and its uncertainty-aware mutants at the level of individual objects. Unlike image-level mutation scores, object-level mutation scores provide a finer-grained view of how predictive uncertainty affects detection outcomes.

We define object-level mutation scores based on the \miss and \ghost sets, respectively. 
Specifically, the miss-based object-level mutation score quantifies the proportion of objects that are correctly detected by the original model but are missed by the mutant models: 
\begin{equation}
    MS_{obj}^{miss} = \frac{|\mathcal{S}_{miss}|}{|\mathcal{S}_{miss}| + |\mathcal{S}_{match}|},
\end{equation}
where $|\mathcal{S}_{miss}|$ and $|\mathcal{S}_{match}|$ denote the number of elements in $\mathcal{S}_{miss}$ and $\mathcal{S}_{match}$, respectively.
\objmsmiss reflects the sensitivity of a test suite to uncertainty-induced false negatives. Higher values indicate that the test suite more effectively exposes scenarios in which injected uncertainty causes the model to lose previously stable detections.

Similarly, the ghost-based object-level mutation score is defined as:
\begin{equation}
    MS_{obj}^{ghost} = \frac{|\mathcal{S}_{ghost}|}{|\mathcal{S}_{miss}| + |\mathcal{S}_{match} + \mathcal{S}_{ghost}|}.
\end{equation}
\objmsghost measures a test suite's ability to kill mutants by exposing false positives, i.e., objects detected by mutant models but not by the original model. Higher values indicate that the test suite is effective at exposing overconfident or unstable predictions, which is particularly critical in safety-sensitive object detection applications.

Finally, we define the combined object-level mutation score, which accounts for both miss and ghost objects, as:
\begin{equation}
    MS_{obj}^{mg}
    =
    \frac{|\mathcal{S}_{miss}| + |\mathcal{S}_{ghost}|}
    {|\mathcal{S}_{miss}| + |\mathcal{S}_{match}| + |\mathcal{S}_{ghost}|}.
\end{equation}
\objmsmg characterizes overall mutant killability at the object level by jointly considering missed and ghost objects. By capturing both the disappearance of originally detected objects and the emergence of additional detections, it provides a comprehensive measure of behavioral divergence introduced by uncertainty-aware mutations.

\subsubsection{IoU-based Mutation Score}\label{subsubsec:iou_ms}
IoU-based mutation score evaluates test suite quality by quantifying degradation in spatial overlap for matched objects detected by both the original model and its mutants, which is defined as:
\begin{equation}
MS_{iou} = \frac{1}{|\mathcal{S}_{match}|} 
\sum_{(d_i, d'_j) \in \mathcal{S}_{match}}
\left( 1 - \mathrm{IoU}(b_i, b'_j) \right),
\end{equation}
where $b_i$, $b'_j$ are the bounding boxes corresponding to the detected objects $d_i$, $d'_j$ from the original model and its mutant, respectively.

\subsubsection{Uncertainty-aware Mutation Score}\label{subsubsec:uncer_ms}
Uncertainty-aware mutation scores evaluate test suite quality by measuring differences in predictive uncertainty between the original model and its mutants. To quantify predictive uncertainty, we execute both the original model and each mutant $n$ times and compute uncertainty metrics from the resulting output distributions. 
Then, we calculate uncertainty-aware mutation scores for each of the match, miss, and ghost object sets.

Specifically, the uncertainty-aware mutation scores are defined as:
\begin{equation}
    UA\text{-}MS = \left| UM_{\text{orig}} - UM_{\text{mut}} \right|,
\end{equation}
where \(UM_{\text{orig}}\) and \(UM_{\text{mut}}\) denote the uncertainty metric computed for the original model and the mutant model, respectively, over a given object set. 
Following Lu et al.~\cite{lu2025assessing} (see Section~\ref{subsec:uncer_dl}), we adopt two categories of uncertainty metrics: for classification tasks, we employ VR, SE, and MI; for regression tasks, we employ TV and PS. 
$UA\text{-}MSs$ quantify the extent to which uncertainty-aware mutations affect the model's predictive uncertainty, thereby reflecting the sensitivity of the test suite to uncertainty-induced behavioral changes. 
We calculate \textit{UA-MS} considering each of the three categories of objects (i.e., $\mathcal{S}_{match}$, $\mathcal{S}_{miss}$, and $\mathcal{S}_{ghost}$) and five uncertainty metrics; therefore, we have 15 (3 object sets $\times$ 5 uncertainty metrics) \textit{UA-MSs} in total.

\section{Experiment Design}

 \subsection{Research Questions}\label{subsec:rqs}
In this paper, we plan to answer the following research questions (RQs).

\begin{itemize}[left=0pt]
    \item RQ1: How do the proposed mutation scores compare with the traditional one? RQ1 compares the proposed mutation scores with the traditional ones regarding their effectiveness in differentiating test suite quality under uncertainty.
    \item RQ2: How well do the proposed mutation scores distinguish test suite quality? RQ2 evaluates the performance of the proposed mutation scores in distinguishing test suites regarding their effectiveness in detecting failures induced by uncertainty. Unlike RQ1, this RQ compares the proposed mutation scores in terms of their ability to differentiate the test suites.
    \item RQ3: How effective are the proposed \glspl{ms} in revealing uncertainty in the model? Based on the assumption that higher model uncertainty leads to a higher failure rate, RQ3 investigated the ability of the proposed \glspl{ms} to detect such degradation under different uncertainty levels. 
\end{itemize}

\subsection{Case Studies}\label{subsec:case_study}
We conduct experiments on three industrial case studies: a laptop sticker–removal robot, a laptop disassembling robot, and an entertainment robot. All three robots are equipped with \gls{dl}-enabled object detection software. The detailed characteristics of each case study are summarized in Table~\ref{tab:case_studies}.

\begin{table}[ht]
    \centering
    \caption{Overview of the industrial case studies and experimental configurations, including subject models, 
    % benchmark datasets, 
    test suites,
    and mutation ratio settings.}
    \resizebox{\textwidth}{!}{\begin{tabular}{llllll}
\toprule
\multirow{1}{*}{Robot} & \multirow{1}{*}{Model} & \multirow{1}{*}{Model Architecture}  & \multirow{1}{*}{Test Suite} & \multicolumn{1}{c}{Mutation Ratio (MCD)} & \multicolumn{1}{c}{Mutation Ratio (MCB)} \\ \midrule
                                % &              &     &      &  &  &     \\ \midrule
\multirow{2}{*}{\er}    &   \hfdm{\textit{1}}         & \textit{YuNet}                  &  \multirow{2}{*}{\lowimg, \mediumimg, \highimg}            & \multirow{8}{*}{
\begin{tabular}{@{}l@{}}
0.10\\
0.15\\
0.20\\
0.25\\
0.30\\
0.35\\
0.40\\
0.45\\
0.50
\end{tabular}}
 &  \multirow{8}{*}{%
 \begin{tabular}{l}
$0.10 \cdot \{1,3,5,7,9\}$\\
$0.15 \cdot \{1,3,5,7,9\}$\\
$0.20 \cdot \{1,3,5,7,9\}$\\
$0.25 \cdot \{1,3,5,7,9\}$\\
$0.30 \cdot \{1,3,5,7,9\}$\\
$0.35 \cdot \{1,3,5,7,9\}$\\
$0.40 \cdot \{1,3,5,7,9\}$\\
$0.45 \cdot \{1,3,5,7,9\}$\\
$0.50 \cdot \{1,3,5,7,9\}$
\end{tabular}
% \begin{tabular}{@{}c@{}}
% (0.10,1) (0.10,3) (0.10,5) (0.10,7) (0.10,9)\\
% (0.15,1) (0.15,3) (0.15,5) (0.15,7) (0.15,9)\\
% (0.20,1) (0.20,3) (0.20,5) (0.20,7) (0.20,9)\\
% (0.25,1) (0.25,3) (0.25,5) (0.25,7) (0.25,9)\\
% (0.30,1) (0.30,3) (0.30,5) (0.30,7) (0.30,9)\\
% (0.35,1) (0.35,3) (0.35,5) (0.35,7) (0.35,9)\\
% (0.40,1) (0.40,3) (0.40,5) (0.40,7) (0.40,9)\\
% (0.45,1) (0.45,3) (0.45,5) (0.45,7) (0.45,9)\\
% (0.50,1) (0.50,3) (0.50,5) (0.50,7) (0.50,9)
% \end{tabular}%
}
             \\
                        & \hfdm{\textit{2}} & \textit{YuNet-s}                &              &                \\ \cmidrule(r){1-4}
\multirow{1}{*}{\ldr}   &   \multirow{1}{*}{\scdm}         & \textit{YOLOv11}                 &  \multirow{1}{*}{\normalimg, \diffexpimg}            &                \\ \cmidrule(r){1-4}
\multirow{5}{*}{\lsrr}   &   \stdm{\textit{1}}     & \fasterrcnn                      & \multirow{5}{*}{\origimg, \dalleimg, \sdimg}             &                \\
                         &   \stdm{\textit{2}} & \fasterrcnnv                     &              &                \\
                         &   \stdm{\textit{3}} & \retinanet                      &              &                \\
                         &   \stdm{\textit{4}} & \retinanetv                     &              &                \\
                         &   \stdm{\textit{5}} & \ssd                 &              &                \\ \bottomrule
\end{tabular}}
    \label{tab:case_studies}
\end{table}

\subsubsection{Laptop Sticker Removing Robot (\lsrr)} 
The Danish Technological Institute (DTI)\footnote{\url{https://www.dti.dk/}} develops robotic technologies that support a sustainable and circular economy. One of its focuses is robotic systems for refurbishing electronic devices, particularly laptops. To support this goal, an \lsrr is developed to automate sticker removal from laptop surfaces, a task that is traditionally manual, time-consuming, and prone to inconsistencies. \lsrr employs a \gls{dl}-enabled sticker detection software (\stdm) to localize stickers and estimate their boundaries. 

\textbf{\textit{Subject \gls{dl}-enabled Software}}. We obtain five pre-trained \stdm{s} employed in \lsrr developed by DTI. These \stdm{s} are based on various object detection architectures, including \fasterrcnn~\cite{ren2015faster}, \fasterrcnnv~\cite{li2021benchmarking}, \retinanet~\cite{lin2017focal}, \retinanetv~\cite{zhang2020bridging}, \ssd~\cite{liu2016ssd}, and \ssdlite~\cite{howard2019searching}. They were trained by DTI using a dedicated sticker detection dataset containing thousands of annotated laptop images with stickers in various poses, sizes, and lighting conditions. All five \stdm{s} are trained based on the open-source PyTorch implementation~\cite{paszke2019pytorch}. 
We modify the five \stdm{s} to be compatible with the two uncertainty-aware mutation operators by injecting prediction-time-activated dropout/dropblock layers. Specifically,  for \fasterrcnn, \fasterrcnnv, \retinanet, and \retinanetv, we add dropout/dropblock layers to the last three convolutional layers of the Feature Pyramid Network, a shared component built on top of their backbones. For \ssd, we add dropout/dropblock layers to the last three layers of the detection head.

\textbf{\textit{Subject Test Suites}}. We select three 
subject test suites:
\origimg, \dalleimg, and \sdimg. \origimg contains real-world images collected by DTI, where stickers are manually placed on laptops and annotated by hand. In contrast, \dalleimg and \sdimg are synthetic test suites generated by prompting two vision–language models to produce realistic laptop images with stickers under diverse conditions. All stickers in \dalleimg and \sdimg are manually annotated, and each image is manually validated to ensure visual realism. Each test suite contains 150 images and has been used in prior studies as a standardized benchmark~\cite{lu2025assessing}.

\subsubsection{Laptop Disassembling Robot (\ldr)} 
\ldr is another robotic system developed by DTI for laptop refurbishment, focusing on automated laptop disassembly. A critical step in the disassembly process is accurate screw localization. To this end, \ldr integrates \gls{dl}-enabled screw detection software (\scdm) to automatically identify screw positions, which is the foundation for precise and fully automated laptop disassembly. 

\textbf{\textit{Subject \gls{dl}-enabled Software}}. We obtain one pretrained \scdm from DTI, which is based on the \textit{YOLOv11}~\cite{varghese2024yolov8} architecture, and we inject prediction-time–activated dropout/dropblock layers into the last three convolutional layers of the \textit{YOLOv11} backbone.

\textbf{\textit{Subject Test Suites}}. We obtain two test suites from DTI, i.e., \normalimg and \diffexpimg, each consisting of images of laptops with screws. \normalimg contains 25 images captured under standard lighting conditions representative of typical operating environments, while \diffexpimg contains 125 images captured under varying exposure levels, including both underexposed and overexposed conditions. Both test suites are manually annotated, providing ground-truth labels for screws and their corresponding screw holders.

\subsubsection{TIAGo Pro} PAL Robotics\footnote{\url{https://pal-robotics.com/}} is a leading robotic manufacturer in Spain, with deployments across diverse domains, including warehouses, retail spaces, healthcare, and offices. 
Notably, \er is an open-source mobile manipulator designed for advanced research and applied development. It integrates autonomous navigation, dexterous manipulation, and multimodal perception within a unified platform, featuring a mobile base, a torque-controlled robotic arm, and configurable end-effectors. 
To operate in dynamic environments, \er incorporates RGB-D cameras, LiDAR, and force–torque sensing, and supports various perception and decision-making modules for tasks such as object detection and human-robot interaction. Specifically, \er employs \gls{dl}-enabled human face detection software (\hfdm) to support its face detection and recognition. 

\textbf{\textit{Subject \gls{dl}-enabled Software}}. We use two pretrained \hfdm{s} in \er, both based on the \textit{YuNet} architecture~\cite{wu2023yunet}. YuNet is a lightweight face detector capable of millisecond-level inference, making it suitable for real-time applications. It is widely applied in embedded systems, mobile devices, and robotic platforms, where efficient and accurate face detection is critical for responsive human-robot interaction. The two \hfdm{s} include the standard \textit{YuNet} and a smaller variant, \textit{YuNet-s}, optimized for faster inference. Similarly, to enable the prediction-time dropout/dropblock, we modify the last three convolutional layers of the common backbones of \textit{YuNet} and \textit{YuNet-s}.

\textbf{\textit{Subject Test Suites}}. According to PAL, the \hfdm{s} are pretrained on public data, and no dedicated test suites were specifically collected for evaluation. Therefore, we adopt WIDER FACE~\cite{yang2016wider}, the largest publicly available face detection dataset, to construct the subject test suites. Concretely, we select the test set of WIDER FACE and partition it into three subsets based on the number of objects per image, using the 25\% (1), 50\% (3), and 75\% (9) percentiles as thresholds. These subsets are denoted as \lowimg, \mediumimg, and \highimg, corresponding to images with low, medium, and high object densities, respectively. Specifically, \lowimg contains 1416 images with 1 to 3 objects, \mediumimg contains 309 images with 4 to 9 objects, and \highimg contains 88 images with 10 or more objects.

\subsection{Mutation Ratio}
Recall from Section~\ref{subsec:mutation_op} that the mutation ratio controls the extent of uncertainty injected by the mutation operators.
In our experiments, we configure the mutation operators with multiple mutation ratios to generate diverse mutants. Table~\ref{tab:case_studies} summarizes the detailed configurations of the mutation ratios. For the \mcd operator, we use nine dropout rates ranging from 0.10 to 0.50 in increments of 0.05 as the mutation ratios. For the \mcb operator, we consider all combinations of these nine dropout rates with five block sizes \{1,3,5,7,9\}, yielding a total of 45 configurations. 
The generated mutants are evaluated across all relevant test suites in each case study. 
These mutation ratios are chosen to systematically explore the models' sensitivity to different levels and patterns of predictive uncertainty, and have also been adopted in prior works~\cite{yelleni2024monte,lu2025assessing,catak2021prediction}.

\subsection{Evaluation Metrics}\label{subsec:metrics}
Table~\ref{tab:exp_design} presents the metrics used to answer each research question (RQ). Since RQ1 focuses on comparing the proposed mutation scores with the traditional mutation score, we use the image-level mutation score (\imgms) as the baseline metric for comparison with the three types of proposed mutation scores (i.e., \objms, \ioums, and \uams). RQ2 and RQ3 focus on evaluating test suite quality using the new metrics and investigating the impact of uncertainty on the mutation scores; therefore, we employ \objms, \ioums, and \uams to perform this analysis.
\begin{table}[ht]
    \centering
    \caption{Description of tasks, employed case studies, metrics, and statistical tests for each RQ.}
    \resizebox{\textwidth}{!}{% \begin{tabular}{llllll}

\begin{tabular}{lllll}
\toprule
RQs      & \begin{tabular}[c]{@{}l@{}}Tasks \\ (Section~\ref{subsec:rqs})\end{tabular}                                                        & \begin{tabular}[c]{@{}l@{}}Case Study \\ (Section~\ref{subsec:case_study})\end{tabular} & \begin{tabular}[c]{@{}l@{}}Metrics \\ (Section~\ref{subsec:metrics})\end{tabular}  & \begin{tabular}[c]{@{}l@{}}Statistical Tests \\ (Section~\ref{subsec:stat_test})\end{tabular} \\ \midrule
RQ1       & Compare the proposed MSs with traditional MSs                                                                                                  & \multirow{4}{*}{\begin{tabular}[c]{@{}l@{}}\er,\\ \ldr,\\ \lsrr\end{tabular}}                   & \begin{tabular}[c]{@{}l@{}}\imgms (baseline metrics),\\ \objms, \ioums, \uams\end{tabular} &    /                                                                                                \\ \cmidrule(r){1-2} \cmidrule(r){4-5}
RQ2       & \begin{tabular}[c]{@{}l@{}}Evaluate how effectively the proposed MSs \\ distinguish the quality of different test suites\end{tabular}          &                                                                                              & \multirow{3}{*}{\objms, \ioums, \uams}                                                     &  
\begin{tabular}[c]{@{}l@{}}Kruskal–Wallis test, \\ Eta-squared (\(\eta ^{2}\)) effect size\end{tabular}     \\ \cmidrule(r){1-2} \cmidrule(r){5-5}
RQ3 & \begin{tabular}[c]{@{}l@{}}Examine how mutant behavior changes under \\ different levels of predictive uncertainty\end{tabular}                &    &   & \begin{tabular}[c]{@{}l@{}}Spearman's correlation test \\ Multiple Correlation Coefficient ($R$), \\F-test \end{tabular}  \\ \bottomrule
                     % & RQ3.2 & \begin{tabular}[c]{@{}l@{}}Assess how different uncertainty levels affect \\ the ability of MSs to distinguish test suite quality\end{tabular} &  &   &  \begin{tabular}[c]{@{}l@{}}Kruskal–Wallis test, \\ Eta-squared (\(\eta ^{2}\)) effect size\end{tabular}       \\ \bottomrule                                                                 
\end{tabular}

}
    \label{tab:exp_design}
\end{table}

\subsection{Statistical Tests}\label{subsec:stat_test}
To answer RQ2, we employ the Kruskal-Wallis rank test~\cite{kruskal1952use} at a significance level of $0.01$ to determine if there are significant differences among all paired groups (test suites), with a \textit{p-value} < 0.01 indicating a significant difference. 
We then calculate eta-squared ($\eta^2$) as the effect size, ranging from 0 to 1. Following commonly used thresholds~\cite{eta_effect_size}, $\eta^2$ values of $0.01 \le \eta^2 < 0.06$ indicate a small effect, $0.06 \le \eta^2 < 0.14$ indicate a moderate effect, and $\eta^2 \ge 0.14$ indicate a large effect.
In RQ3, we study the correlation between mutation ratios and mutation scores to see how the change of uncertainty affects the performance of mutation scores. Therefore, we perform the Spearman's rank correlation ($\rho$) test~\cite{zar2005spearman}, a non-parametric test that measures the monotonic relationship between two ranked variables. Values of $\rho \in (0, 1]$ indicate a positive correlation, $\rho \in [-1, 0)$ shows a negative correlation, 1.0 (-1.0) indicates a perfect positive (negative) correlation, and 0 means no correlation. Correlation is considered statistically significant if \textit{p-value} < 0.01. Following Mukaka~\cite{mukaka2012guide}, we divide $\rho$ into five levels: \textit{negligible} ($\rho$ $\in$ (-0.300, 0.300)), \textit{low} ($\rho$ $\in$ [0.300, 0.500) or (-0.500, -0.300]), \textit{moderate} ($\rho$ $\in$ [0.50, 0.700) or (-0.700, 0.500]), \textit{high} ($\rho$ $\in$ [0.700, 0.900) or (-0.900, -0.700]), and \textit{very high} ($\rho$ $\in$ [0.900, 1.000] or [-1.000, -0.900]). Recall that \mcb operator has two mutation ratios (dropout rate and block size) that together control the mutation of a model; therefore, for \mcb, we employ the multiple correlation coefficient ($R$)~\cite{asuero2006correlation} to measure the effect of the combined mutation ratios on mutation scores. 
F-test is applied to assess the statistical significance of the multiple correlation test, with \textit{p-value} < 0.01 indicating a significant correlation.
Following Hinkle et al.~\cite{hinkle2003applied}, $R$ is interpreted using five levels: \textit{negligible} ($R$ $\in$ [0, 0.300)), \textit{low} ($R$ $\in$ [0.300, 0.500), \textit{moderate} ($R$ $\in$ [0.50, 0.700), \textit{high} ($R$ $\in$ [0.700, 0.900), and \textit{very high} ($R$ $\in$ [0.900, 1.000].

\section{Results and Analysis}

\subsection{Results for RQ1 -- Proposed vs. Traditional Mutation Scores}
Table~\ref{tab:avg_ms_1} and Table~\ref{tab:avg_ms_2} present the descriptive results of mutation scores achieved by each test suite across different case studies. For the \er case study, which has two subject models (\hfdm{1} and \hfdm{2}) and three subject test suites (\lowimg, \mediumimg, and \highimg), all three \imgms metrics ( \imgmsmiss, \imgmsghost, and \imgmsmg) do not differentiate among the test suites. In contrast, the three \objms and the \ioums metrics exhibit a clear and consistent decreasing trend from \lowimg to \highimg for both models, differentiating the three \glspl{ts}. Similarly, the \uams metrics largely follow this pattern: 11 out of the 15 \uams metrics demonstrate a clear decreasing trend from \lowimg to \mediumimg and further to \highimg, indicating that test suites with fewer objects are more effective at revealing uncertainty-induced failures.
\begin{table}[ht]
    \centering
    \caption{Mutation Scores Achieved by Different Test Suites on \hfdm{1}, \hfdm{2}, \scdm{1}, and \stdm{1} -- RQ1.}
    \resizebox{\textwidth}{!}{\begin{tabular}{llccc ccc cc ccc}
\toprule
\multicolumn{2}{c}{\multirow{2}{*}{\textit{Metrics}}} 
& \multicolumn{3}{c}{\hfdm{1}}
& \multicolumn{3}{c}{\hfdm{2}}
& \multicolumn{2}{c}{\scdm{1}}
& \multicolumn{3}{c}{\stdm{1}} \\
\cmidrule(lr){3-5} \cmidrule(lr){6-8} \cmidrule(lr){9-10} \cmidrule(lr){11-13}
 & & \lowimg & \mediumimg & \highimg
 & \lowimg & \mediumimg & \highimg
 & \normalimg & \diffexpimg
 & \origimg & \dalleimg & \sdimg \\
\midrule
\multirow{3}{*}{\textit{Img-MS}} 
& $MS_{img}^{miss}$  & 1.000 & 1.000 & 1.000 & 1.000 & 1.000 & 1.000 & 0.000 & 1.000 & 0.481 & 1.000 & 1.000 \\
& $MS_{img}^{ghost}$ & 1.000 & 1.000 & 1.000 & 1.000 & 1.000 & 1.000 & 0.204 & 0.796 & 1.000 & 1.000 & 1.000 \\
& $MS_{img}^{mg}$    & 1.000 & 1.000 & 1.000 & 1.000 & 1.000 & 1.000 & 0.204 & 1.000 & 1.000 & 1.000 & 1.000 \\
\midrule
\multirow{3}{*}{\textit{Obj-MS}} 
& $MS_{obj}^{miss}$  & 0.499 & 0.320 & 0.192 & 0.482 & 0.273 & 0.165 & 0.000 & 0.006 & 0.000 & 0.015 & 0.025 \\
& $MS_{obj}^{ghost}$ & 0.346 & 0.270 & 0.175 & 0.461 & 0.337 & 0.217 & 0.000 & 0.002 & 0.009 & 0.014 & 0.019 \\
& $MS_{obj}^{mg}$    & 0.662 & 0.496 & 0.327 & 0.708 & 0.507 & 0.336 & 0.000 & 0.008 & 0.010 & 0.028 & 0.043 \\
\midrule
\multirow{1}{*}{\textit{IoU-MS}} 
& $MS_{iou}$ & 0.580 & 0.419 & 0.304 & 0.563 & 0.378 & 0.270 & 0.034 & 0.048 & 0.049 & 0.065 & 0.072 \\
\midrule
\multirow{15}{*}{\textit{UA-MS}} 
& $MS_{vr}^{match}$ & 0.476 & 0.317 & 0.208 & 0.462 & 0.278 & 0.180 & 0.000 & 0.006 & 0.000 & 0.015 & 0.023 \\
& $MS_{ie}^{match}$ & 0.362 & 0.239 & 0.152 & 0.356 & 0.210 & 0.130 & 0.003 & 0.014 & 0.007 & 0.020 & 0.026 \\
& $MS_{mi}^{match}$ & 0.476 & 0.309 & 0.194 & 0.461 & 0.270 & 0.169 & 0.000 & 0.007 & 0.001 & 0.018 & 0.028 \\
& $MS_{va}^{match}$ & 0.974 & 0.963 & 0.929 & 0.966 & 0.951 & 0.899 & 0.410 & 0.407 & 0.900 & 0.969 & 0.972 \\
& $MS_{ps}^{match}$ & 0.957 & 0.971 & 0.975 & 0.958 & 0.973 & 0.975 & 0.911 & 0.910 & 0.984 & 0.991 & 0.992 \\ \cmidrule(lr){2-13}
& $MS_{vr}^{miss}$ & 0.606 & 0.539 & 0.528 & 0.593 & 0.507 & 0.496 & 0.000 & 0.015 & 0.001 & 0.063 & 0.071 \\
& $MS_{ie}^{miss}$ & 0.469 & 0.418 & 0.415 & 0.460 & 0.394 & 0.386 & 0.000 & 0.014 & 0.001 & 0.036 & 0.040 \\
& $MS_{mi}^{miss}$ & 0.606 & 0.539 & 0.528 & 0.593 & 0.507 & 0.496 & 0.000 & 0.015 & 0.001 & 0.063 & 0.071 \\
& $MS_{va}^{miss}$ & 0.606 & 0.539 & 0.528 & 0.593 & 0.507 & 0.496 & 0.000 & 0.015 & 0.001 & 0.063 & 0.071 \\
& $MS_{ps}^{miss}$ & 0.606 & 0.539 & 0.528 & 0.593 & 0.507 & 0.496 & 0.000 & 0.015 & 0.001 & 0.063 & 0.071 \\ \cmidrule(lr){2-13}
& $MS_{vr}^{ghost}$ & 0.154 & 0.153 & 0.146 & 0.152 & 0.152 & 0.147 & 0.001 & 0.005 & 0.020 & 0.046 & 0.060 \\
& $MS_{ie}^{ghost}$ & 0.101 & 0.101 & 0.099 & 0.100 & 0.100 & 0.099 & 0.001 & 0.005 & 0.007 & 0.016 & 0.020 \\
& $MS_{mi}^{ghost}$ & 0.152 & 0.152 & 0.145 & 0.151 & 0.151 & 0.147 & 0.001 & 0.005 & 0.019 & 0.044 & 0.058 \\
& $MS_{va}^{ghost}$ & 0.076 & 0.077 & 0.080 & 0.081 & 0.082 & 0.083 & 0.001 & 0.004 & 0.007 & 0.013 & 0.015 \\
& $MS_{ps}^{ghost}$ & 0.102 & 0.103 & 0.104 & 0.102 & 0.102 & 0.103 & 0.001 & 0.004 & 0.010 & 0.020 & 0.027 \\
\bottomrule
\end{tabular}
}
    \label{tab:avg_ms_1}
    \vspace{2pt}
\end{table}
\begin{table}[ht]
    \centering
    \caption{Mutation Scores Achieved by Different Test Suites on \stdm{2}, \stdm{3}, \stdm{4}, \stdm{5} -- RQ1.}
    \resizebox{\textwidth}{!}{\begin{tabular}{ll rrr rrr rrr rrr}
\toprule
\multicolumn{2}{c}{\multirow{2}{*}{\textit{Metrics}}} 
& \multicolumn{3}{c}{\stdm{2}} & \multicolumn{3}{c}{\stdm{3}} & \multicolumn{3}{c}{\stdm{4}} & \multicolumn{3}{c}{\stdm{5}} \\
\cmidrule(lr){3-5} \cmidrule(lr){6-8} \cmidrule(lr){9-11} \cmidrule(lr){12-14}
 % & & orig & dalle & sd & orig & dalle & sd & orig & dalle & sd & orig & dalle & sd \\
 & & \origimg & \dalleimg & \sdimg
 & \origimg & \dalleimg & \sdimg
 & \origimg & \dalleimg & \sdimg
 & \origimg & \dalleimg & \sdimg \\
\midrule
\multirow{3}{*}{\textit{Img-MS}} 
& $MS_{img}^{miss}$ & 0.593 & 1.000 & 1.000 & 1.000 & 1.000 & 1.000 & 0.981 & 0.778 & 1.000 & 1.000 & 1.000 & 1.000 \\
& $MS_{img}^{ghost}$ & 0.889 & 1.000 & 1.000 & 1.000 & 1.000 & 1.000 & 1.000 & 1.000 & 1.000 & 1.000 & 1.000 & 1.000 \\
& $MS_{img}^{mg}$ & 0.907 & 1.000 & 1.000 & 1.000 & 1.000 & 1.000 & 1.000 & 1.000 & 1.000 & 1.000 & 1.000 & 1.000 \\
\midrule
\multirow{3}{*}{\textit{Obj-MS}} 
& $MS_{obj}^{miss}$ & 0.002 & 0.029 & 0.037 & 0.010 & 0.009 & 0.022 & 0.005 & 0.006 & 0.025 & 0.274 & 0.049 & 0.049 \\
& $MS_{obj}^{ghost}$ & 0.004 & 0.012 & 0.032 & 0.006 & 0.012 & 0.006 & 0.002 & 0.005 & 0.010 & 0.523 & 0.474 & 0.490 \\
& $MS_{obj}^{mg}$ & 0.005 & 0.041 & 0.067 & 0.015 & 0.021 & 0.027 & 0.007 & 0.011 & 0.035 & 0.635 & 0.500 & 0.515 \\
\midrule
\multirow{1}{*}{\textit{Obj-MS}} 
& $MS_{iou}$ & 0.044 & 0.067 & 0.070 & 0.044 & 0.042 & 0.052 & 0.034 & 0.029 & 0.047 & 0.389 & 0.262 & 0.272 \\ \midrule
\multirow{15}{*}{\textit{UA-MS}} 
& $MS_{vr}^{match}$ & 0.002 & 0.027 & 0.031 & 0.010 & 0.008 & 0.021 & 0.005 & 0.006 & 0.027 & 0.181 & 0.049 & 0.049 \\
& $MS_{ie}^{match}$ & 0.020 & 0.043 & 0.048 & 0.020 & 0.021 & 0.032 & 0.009 & 0.015 & 0.025 & 0.211 & 0.098 & 0.089 \\
& $MS_{mi}^{match}$ & 0.005 & 0.035 & 0.039 & 0.013 & 0.012 & 0.025 & 0.005 & 0.009 & 0.027 & 0.202 & 0.073 & 0.069 \\
& $MS_{va}^{match}$ & 0.889 & 0.967 & 0.971 & 0.838 & 0.941 & 0.952 & 0.744 & 0.815 & 0.778 & 0.949 & 0.999 & 0.999 \\
& $MS_{ps}^{match}$ & 0.983 & 0.985 & 0.991 & 0.979 & 0.984 & 0.989 & 0.966 & 0.942 & 0.911 & 0.948 & 0.998 & 0.998 \\ \cmidrule(lr){2-14}
& $MS_{vr}^{miss}$ & 0.004 & 0.089 & 0.098 & 0.026 & 0.023 & 0.062 & 0.013 & 0.013 & 0.070 & 0.345 & 0.088 & 0.097 \\
& $MS_{ie}^{miss}$ & 0.004 & 0.053 & 0.067 & 0.026 & 0.022 & 0.054 & 0.013 & 0.012 & 0.054 & 0.305 & 0.079 & 0.086 \\
& $MS_{mi}^{miss}$ & 0.004 & 0.089 & 0.098 & 0.026 & 0.023 & 0.062 & 0.013 & 0.013 & 0.070 & 0.345 & 0.088 & 0.097 \\
& $MS_{va}^{miss}$ & 0.004 & 0.089 & 0.098 & 0.026 & 0.023 & 0.062 & 0.013 & 0.013 & 0.070 & 0.345 & 0.088 & 0.097 \\
& $MS_{ps}^{miss}$ & 0.004 & 0.089 & 0.098 & 0.026 & 0.023 & 0.062 & 0.013 & 0.013 & 0.070 & 0.345 & 0.088 & 0.097 \\ \cmidrule(lr){2-14}
& $MS_{vr}^{ghost}$ & 0.011 & 0.036 & 0.094 & 0.015 & 0.048 & 0.023 & 0.005 & 0.013 & 0.037 & 0.199 & 0.214 & 0.216 \\
& $MS_{ie}^{ghost}$ & 0.005 & 0.016 & 0.029 & 0.007 & 0.041 & 0.022 & 0.003 & 0.012 & 0.026 & 0.133 & 0.133 & 0.135 \\
& $MS_{mi}^{ghost}$ & 0.011 & 0.034 & 0.089 & 0.016 & 0.048 & 0.023 & 0.005 & 0.013 & 0.036 & 0.192 & 0.205 & 0.207 \\
& $MS_{va}^{ghost}$ & 0.006 & 0.014 & 0.019 & 0.006 & 0.010 & 0.012 & 0.002 & 0.006 & 0.008 & 0.060 & 0.056 & 0.056 \\
& $MS_{ps}^{ghost}$ & 0.009 & 0.021 & 0.031 & 0.007 & 0.014 & 0.018 & 0.003 & 0.007 & 0.011 & 0.110 & 0.107 & 0.107 \\
\bottomrule
\end{tabular}
}
    \label{tab:avg_ms_2}
\end{table}

In terms of the \ldr case study, which has one subject model (\stdm{1}) and two test suites (\normalimg and \diffexpimg), all \imgms, \objms, and \ioums metrics show clear differences between the two test suites, and 13 out of 15 \uams yield a higher quality of \diffexpimg than \normalimg, suggesting that \diffexpimg is more effective in exposing uncertainty-related failures for \stdm{1}. 

The \lsrr case study has five subject models (\stdm{1} -- \stdm{5}) and three test suites (\origimg, \dalleimg, and \sdimg). For \stdm{1}, the \imgmsmiss metric indicates that \origimg has the worst quality, but cannot distinguish between \dalleimg and \sdimg. For \imgmsghost and \imgmsmg, they remain consistently the same across all three test suites, indicating limited distinguishing ability. In contrast, the \objms and \ioums metrics exhibit an increasing trend from \origimg to \dalleimg and further to \sdimg, indicating that these metrics are more sensitive in differentiating test suites. Similarly, all \uams metrics consistently reveal improved quality from \origimg, \dalleimg, to \sdimg. This pattern extends to the remaining models (\stdm{2}–\stdm{5}), where \imgms metrics remain identical across all test suites in 8 out of 12 cases, whereas \objms, \ioums, and \uams consistently reveal differences in test suites. Overall, these results indicate that proposed mutation scores (\objms, \ioums, and \uams) are consistently more effective than \imgms in distinguishing the quality of different test suites.

\begin{center}
    \fcolorbox{black}{gray!10}{\parbox{\textwidth}{\textbf{Conclusion for 
    RQ1}: Compared to the traditional mutation score (\imgms), the novel mutation scores (\objms, \ioums, and \uams) are more effective at distinguishing test suite quality in revealing uncertainty-induced failures in \gls{dl}-enabled software across six out of seven models in three case studies. This indicates that the proposed mutation score metrics provide a more sensitive and reliable assessment of test suite quality than traditional ones in the presence of uncertainty.
    }}
\end{center}

\subsection{Results for RQ2 -- Effectiveness in Distinguishing Test Suites}
To answer RQ2, we apply the non-parametric Kruskal–Wallis test to compute $p$-\textit{values}, assessing whether statistically significant differences exist among test suites. We also calculate eta-squared ($\eta^2$) to quantify the effect size. The results are shown in Table~\ref{tab:global_stats}. 
For \objms metrics, \objmsmiss and \objmsmg achieve statistically significant results on all models ($p < 0.01$), indicating their strong performance in distinguishing test suite quality. Notably, \objmsmiss consistently achieves a large effect size ($\eta^2 \ge 0.14$) across all models, whereas \objmsmg shows large effect sizes for the majority but moderate effect sizes for \stdm{3} and \stdm{5}. The performance of \objmsghost is also strong with significant differences and large effect sizes for most models, except for \stdm{5}, where the difference is not statistically significant ($p = 0.271$). 
\begin{table}[ht]
    \centering
    \caption{\textbf{Statistical Results Using the Kruskal–Wallis Test and Eta-Squared Effect Size -- RQ2.} A bold $\eta^2$ with $p < 0.01$ indicates a statistically significant difference. The $\eta^2$ effect sizes are further interpreted as: \colorbox{corrVeryHigh}{Large} if $\eta^2 \ge 0.14$, \colorbox{corrHigh}{Moderate} if $0.06 \le \eta^2 < 0.14$, and \colorbox{corrModerate}{Small} if $0.01 \le \eta^2 < 0.06$.
    }
    \resizebox{\textwidth}{!}{% \begin{tabular}{lllllllllllllllll}

\begin{tabular}{llllllllllllllllll}
\toprule
\multicolumn{2}{c}{\multirow{2}{*}{\textit{Metrics}}} & \multicolumn{2}{c}{\hfdm{1}} & \multicolumn{2}{c}{\hfdm{2}} & \multicolumn{2}{c}{\scdm{1}} & \multicolumn{2}{c}{\stdm{1}} & \multicolumn{2}{c}{\stdm{2}} & \multicolumn{2}{c}{\stdm{3}} & \multicolumn{2}{c}{\stdm{4}} & \multicolumn{2}{c}{\stdm{5}} \\
 & & $p$ & $\eta^2$ & $p$ & $\eta^2$ & $p$ & $\eta^2$ & $p$ & $\eta^2$ & $p$ & $\eta^2$ & $p$ & $\eta^2$ & $p$ & $\eta^2$ & $p$ & $\eta^2$ \\
\midrule
\multirow{3}{*}{\textit{Obj-MS}} & $MS_{obj}^{miss}$ & \textbf{<0.01} & \cellcolor{corrVeryHigh}\textbf{0.604} & \textbf{<0.01} & \cellcolor{corrVeryHigh}\textbf{0.692} & \textbf{<0.01} & \cellcolor{corrVeryHigh}\textbf{0.856} & \textbf{<0.01} & \cellcolor{corrVeryHigh}\textbf{0.747} & \textbf{<0.01} & \cellcolor{corrVeryHigh}\textbf{0.653} & \textbf{<0.01} & \cellcolor{corrVeryHigh}\textbf{0.385} & \textbf{<0.01} & \cellcolor{corrVeryHigh}\textbf{0.565} & \textbf{<0.01} & \cellcolor{corrVeryHigh}\textbf{0.571} \\
 & $MS_{obj}^{ghost}$ & \textbf{<0.01} & \cellcolor{corrVeryHigh}\textbf{0.433} & \textbf{<0.01} & \cellcolor{corrVeryHigh}\textbf{0.384} & \textbf{<0.01} & \cellcolor{corrVeryHigh}\textbf{0.276} & \textbf{<0.01} & \cellcolor{corrVeryHigh}\textbf{0.236} & \textbf{<0.01} & \cellcolor{corrVeryHigh}\textbf{0.730} & \textbf{<0.01} & \cellcolor{corrVeryHigh}\textbf{0.370} & \textbf{<0.01} & \cellcolor{corrVeryHigh}\textbf{0.664} & 0.271 & 0.004 \\
 & $MS_{obj}^{mg}$ & \textbf{<0.01} & \cellcolor{corrVeryHigh}\textbf{0.545} & \textbf{<0.01} & \cellcolor{corrVeryHigh}\textbf{0.544} & \textbf{<0.01} & \cellcolor{corrVeryHigh}\textbf{0.717} & \textbf{<0.01} & \cellcolor{corrVeryHigh}\textbf{0.612} & \textbf{<0.01} & \cellcolor{corrVeryHigh}\textbf{0.703} & \textbf{<0.01} & \cellcolor{corrHigh}\textbf{0.113} & \textbf{<0.01} & \cellcolor{corrVeryHigh}\textbf{0.625} & \textbf{<0.01} & \cellcolor{corrHigh}\textbf{0.106} \\ \midrule
\textit{IoU-MS} & $MS_{iou}$ & \textbf{<0.01} & \cellcolor{corrVeryHigh}\textbf{0.639} & \textbf{<0.01} & \cellcolor{corrVeryHigh}\textbf{0.712} & \textbf{<0.01} & \cellcolor{corrHigh}\textbf{0.132} & \textbf{<0.01} & \cellcolor{corrVeryHigh}\textbf{0.185} & \textbf{<0.01} & \cellcolor{corrVeryHigh}\textbf{0.310} & \textbf{<0.01} & \cellcolor{corrModerate}\textbf{0.054} & \textbf{<0.01} & \cellcolor{corrVeryHigh}\textbf{0.212} & \textbf{<0.01} & \cellcolor{corrVeryHigh}\textbf{0.294} \\ \midrule
\multirow{15}{*}{\textit{UA-MS}} & $MS_{vr}^{match}$ & \textbf{<0.01} & \cellcolor{corrVeryHigh}\textbf{0.559} & \textbf{<0.01} & \cellcolor{corrVeryHigh}\textbf{0.642} & \textbf{<0.01} & \cellcolor{corrVeryHigh}\textbf{0.856} & \textbf{<0.01} & \cellcolor{corrVeryHigh}\textbf{0.746} & \textbf{<0.01} & \cellcolor{corrVeryHigh}\textbf{0.639} & \textbf{<0.01} & \cellcolor{corrVeryHigh}\textbf{0.439} & \textbf{<0.01} & \cellcolor{corrVeryHigh}\textbf{0.596} & \textbf{<0.01} & \cellcolor{corrVeryHigh}\textbf{0.551} \\
 & $MS_{ie}^{match}$ & \textbf{<0.01} & \cellcolor{corrVeryHigh}\textbf{0.703} & \textbf{<0.01} & \cellcolor{corrVeryHigh}\textbf{0.769} & \textbf{<0.01} & \cellcolor{corrVeryHigh}\textbf{0.516} & \textbf{<0.01} & \cellcolor{corrVeryHigh}\textbf{0.421} & \textbf{<0.01} & \cellcolor{corrVeryHigh}\textbf{0.252} & \textbf{<0.01} & \cellcolor{corrHigh}\textbf{0.136} & \textbf{<0.01} & \cellcolor{corrVeryHigh}\textbf{0.348} & \textbf{<0.01} & \cellcolor{corrVeryHigh}\textbf{0.423} \\
 & $MS_{mi}^{match}$ & \textbf{<0.01} & \cellcolor{corrVeryHigh}\textbf{0.625} & \textbf{<0.01} & \cellcolor{corrVeryHigh}\textbf{0.694} & \textbf{<0.01} & \cellcolor{corrVeryHigh}\textbf{0.743} & \textbf{<0.01} & \cellcolor{corrVeryHigh}\textbf{0.747} & \textbf{<0.01} & \cellcolor{corrVeryHigh}\textbf{0.572} & \textbf{<0.01} & \cellcolor{corrVeryHigh}\textbf{0.332} & \textbf{<0.01} & \cellcolor{corrVeryHigh}\textbf{0.568} & \textbf{<0.01} & \cellcolor{corrVeryHigh}\textbf{0.513} \\
 & $MS_{va}^{match}$ & \textbf{<0.01} & \cellcolor{corrVeryHigh}\textbf{0.572} & \textbf{<0.01} & \cellcolor{corrVeryHigh}\textbf{0.588} & 0.902 & 0.000 & \textbf{<0.01} & \cellcolor{corrVeryHigh}\textbf{0.473} & \textbf{<0.01} & \cellcolor{corrVeryHigh}\textbf{0.538} & \textbf{<0.01} & \cellcolor{corrVeryHigh}\textbf{0.486} & \textbf{<0.01} & \cellcolor{corrVeryHigh}\textbf{0.145} & \textbf{<0.01} & \cellcolor{corrVeryHigh}\textbf{0.651} \\
 & $MS_{ps}^{match}$ & \textbf{<0.01} & \cellcolor{corrVeryHigh}\textbf{0.226} & \textbf{<0.01} & \cellcolor{corrVeryHigh}\textbf{0.307} & 0.888 & 0.000 & \textbf{<0.01} & \cellcolor{corrVeryHigh}\textbf{0.430} & \textbf{<0.01} & \cellcolor{corrVeryHigh}\textbf{0.545} & \textbf{<0.01} & \cellcolor{corrVeryHigh}\textbf{0.402} & \textbf{<0.01} & \cellcolor{corrVeryHigh}\textbf{0.855} & \textbf{<0.01} & \cellcolor{corrVeryHigh}\textbf{0.626} \\ \cmidrule(lr){2-18}
 & $MS_{vr}^{miss}$ & \textbf{<0.01} & \cellcolor{corrHigh}\textbf{0.107} & \textbf{<0.01} & \cellcolor{corrVeryHigh}\textbf{0.203} & \textbf{<0.01} & \cellcolor{corrVeryHigh}\textbf{0.857} & \textbf{<0.01} & \cellcolor{corrVeryHigh}\textbf{0.673} & \textbf{<0.01} & \cellcolor{corrVeryHigh}\textbf{0.661} & \textbf{<0.01} & \cellcolor{corrVeryHigh}\textbf{0.383} & \textbf{<0.01} & \cellcolor{corrVeryHigh}\textbf{0.600} & \textbf{<0.01} & \cellcolor{corrVeryHigh}\textbf{0.557} \\
 & $MS_{ie}^{miss}$ & \textbf{<0.01} & \cellcolor{corrHigh}\textbf{0.123} & \textbf{<0.01} & \cellcolor{corrVeryHigh}\textbf{0.247} & \textbf{<0.01} & \cellcolor{corrVeryHigh}\textbf{0.856} & \textbf{<0.01} & \cellcolor{corrVeryHigh}\textbf{0.670} & \textbf{<0.01} & \cellcolor{corrVeryHigh}\textbf{0.638} & \textbf{<0.01} & \cellcolor{corrVeryHigh}\textbf{0.362} & \textbf{<0.01} & \cellcolor{corrVeryHigh}\textbf{0.578} & \textbf{<0.01} & \cellcolor{corrVeryHigh}\textbf{0.534} \\
 & $MS_{mi}^{miss}$ & \textbf{<0.01} & \cellcolor{corrHigh}\textbf{0.107} & \textbf{<0.01} & \cellcolor{corrVeryHigh}\textbf{0.203} & \textbf{<0.01} & \cellcolor{corrVeryHigh}\textbf{0.856} & \textbf{<0.01} & \cellcolor{corrVeryHigh}\textbf{0.673} & \textbf{<0.01} & \cellcolor{corrVeryHigh}\textbf{0.661} & \textbf{<0.01} & \cellcolor{corrVeryHigh}\textbf{0.383} & \textbf{<0.01} & \cellcolor{corrVeryHigh}\textbf{0.600} & \textbf{<0.01} & \cellcolor{corrVeryHigh}\textbf{0.557} \\
 & $MS_{va}^{miss}$ & \textbf{<0.01} & \cellcolor{corrHigh}\textbf{0.107} & \textbf{<0.01} & \cellcolor{corrVeryHigh}\textbf{0.203} & \textbf{<0.01} & \cellcolor{corrVeryHigh}\textbf{0.857} & \textbf{<0.01} & \cellcolor{corrVeryHigh}\textbf{0.673} & \textbf{<0.01} & \cellcolor{corrVeryHigh}\textbf{0.661} & \textbf{<0.01} & \cellcolor{corrVeryHigh}\textbf{0.383} & \textbf{<0.01} & \cellcolor{corrVeryHigh}\textbf{0.600} & \textbf{<0.01} & \cellcolor{corrVeryHigh}\textbf{0.557} \\
 & $MS_{ps}^{miss}$ & \textbf{<0.01} & \cellcolor{corrHigh}\textbf{0.107} & \textbf{<0.01} & \cellcolor{corrVeryHigh}\textbf{0.203} & \textbf{<0.01} & \cellcolor{corrVeryHigh}\textbf{0.857} & \textbf{<0.01} & \cellcolor{corrVeryHigh}\textbf{0.673} & \textbf{<0.01} & \cellcolor{corrVeryHigh}\textbf{0.661} & \textbf{<0.01} & \cellcolor{corrVeryHigh}\textbf{0.383} & \textbf{<0.01} & \cellcolor{corrVeryHigh}\textbf{0.600} & \textbf{<0.01} & \cellcolor{corrVeryHigh}\textbf{0.557} \\ \cmidrule(lr){2-18}
 & $MS_{vr}^{ghost}$ & \textbf{<0.01} & \cellcolor{corrModerate}\textbf{0.050} & 0.040 & 0.028 & \textbf{<0.01} & \cellcolor{corrVeryHigh}\textbf{0.266} & \textbf{<0.01} & \cellcolor{corrVeryHigh}\textbf{0.521} & \textbf{<0.01} & \cellcolor{corrVeryHigh}\textbf{0.785} & \textbf{<0.01} & \cellcolor{corrVeryHigh}\textbf{0.509} & \textbf{<0.01} & \cellcolor{corrVeryHigh}\textbf{0.783} & 0.020 & 0.036 \\
 & $MS_{ie}^{ghost}$ & 0.340 & 0.001 & 0.400 & 0.000 & \textbf{<0.01} & \cellcolor{corrVeryHigh}\textbf{0.330} & \textbf{<0.01} & \cellcolor{corrVeryHigh}\textbf{0.415} & \textbf{<0.01} & \cellcolor{corrVeryHigh}\textbf{0.696} & \textbf{<0.01} & \cellcolor{corrVeryHigh}\textbf{0.653} & \textbf{<0.01} & \cellcolor{corrVeryHigh}\textbf{0.804} & 0.536 & 0.000 \\
 & $MS_{mi}^{ghost}$ & \textbf{<0.01} & \cellcolor{corrModerate}\textbf{0.047} & 0.048 & 0.026 & \textbf{<0.01} & \cellcolor{corrVeryHigh}\textbf{0.266} & \textbf{<0.01} & \cellcolor{corrVeryHigh}\textbf{0.525} & \textbf{<0.01} & \cellcolor{corrVeryHigh}\textbf{0.787} & \textbf{<0.01} & \cellcolor{corrVeryHigh}\textbf{0.494} & \textbf{<0.01} & \cellcolor{corrVeryHigh}\textbf{0.774} & 0.029 & 0.032 \\
 & $MS_{va}^{ghost}$ & \textbf{<0.01} & \cellcolor{corrHigh}\textbf{0.087} & 0.183 & 0.009 & \textbf{<0.01} & \cellcolor{corrVeryHigh}\textbf{0.236} & \textbf{<0.01} & \cellcolor{corrVeryHigh}\textbf{0.257} & \textbf{<0.01} & \cellcolor{corrVeryHigh}\textbf{0.428} & \textbf{<0.01} & \cellcolor{corrVeryHigh}\textbf{0.193} & \textbf{<0.01} & \cellcolor{corrVeryHigh}\textbf{0.484} & \textbf{<0.01} & \cellcolor{corrModerate}\textbf{0.056} \\
 & $MS_{ps}^{ghost}$ & \textbf{<0.01} & \cellcolor{corrVeryHigh}\textbf{0.177} & \textbf{<0.01} & \cellcolor{corrVeryHigh}\textbf{0.171} & \textbf{<0.01} & \cellcolor{corrVeryHigh}\textbf{0.251} & \textbf{<0.01} & \cellcolor{corrVeryHigh}\textbf{0.330} & \textbf{<0.01} & \cellcolor{corrVeryHigh}\textbf{0.500} & \textbf{<0.01} & \cellcolor{corrVeryHigh}\textbf{0.268} & \textbf{<0.01} & \cellcolor{corrVeryHigh}\textbf{0.347} & \textbf{<0.01} & \cellcolor{corrHigh}\textbf{0.095} \\
\bottomrule
\end{tabular}}
    \label{tab:global_stats}
    % \vspace{-10pt}
\end{table}
Similarly, the \ioums metric can significantly distinguish between different test suites for all models, and the effect sizes vary from large magnitudes for \hfdm{2} ($\eta^2 = 0.712$) to moderate for \scdm{1} ($\eta^2 = 0.132$) and small for \stdm{3} ($\eta^2 = 0.054$). 

Regarding the \uams metrics, the ability to distinguish test suite quality varies across metrics. When looking at \uams calculated based on the match set ($\mathcal{S}_{match}$), \uamsmatchvr, \uamsmatchie, and \uamsmatchmi demonstrate consistently significant differences among different test suites with large effect sizes across nearly all models, indicating that these metrics are highly effective at distinguishing test suite quality. \uamsmatchva and \uamsmatchps also demonstrate significant differences among different test suites with large effect sizes on six out of seven models, with one exception observed for \scdm{1}, where the results are not statistically significant ($p > 0.01$). 
For \uams metrics calculated based on $\mathcal{S}_{miss}$ set (\uamsmissvr, \uamsmissie, \uamsmissmi, \uamsmissva, and \uamsmissps), a stable pattern is observed where all metrics consistently yield statistically significant differences among different test suites across all models; the effect sizes are large for six out of seven models, with \hfdm{1} exhibiting moderate magnitudes. 
Finally, the performance of the ghost set ($\mathcal{S}_{ghost}$) based \uams metrics varies across models. \uamsghostvr, \uamsghostie, \uamsghostmi, and \uamsghostva are highly effective at distinguishing test suites for \scdm{1} and \stdm{1}--\stdm{4}, with significant differences and large effect sizes observed; however, for \hfdm{1}, \hfdm{2}, and \stdm{5}, these metrics exhibit non-significant or significant but small effect sizes. One possible explanation for this discrepancy is that the mutants associated with models such as \hfdm detect fewer ghost objects. Consequently, the limited number of detected ghost objects results in lower variance within these metrics, reducing their ability to distinguish between test suites.
Noteably, \uamsghostps demonstrates statistically significant differences among different test suites for all models, with effect sizes ranging from moderate to large.

\begin{center}
    \fcolorbox{black}{gray!10}{\parbox{\textwidth}{\textbf{Conclusion for 
    RQ2}: The proposed mutation scores are effective in distinguishing the test suite's ability to reveal uncertainty-induced failures in \gls{dl}-enabled software. 
    In particular, \ioums and \objms (especially \objmsmiss and \objmsmg) consistently demonstrate statistically significant differences in test suites across all models. Among \uams, \uamsmatchvr, \uamsmatchie, \uamsmatchmi, all \uams derived from $\mathcal{S}_{miss}$, and \uamsghostps are the most stable and effective in highlighting differences between test suites, indicating their sensitivity to uncertainty-induced failures.
    }}
\end{center}

\subsection{Results for RQ3 -- Effect of Uncertainty on Mutation Scores}
To answer RQ3, we first visualize the trends of each mutation score as the mutation ratio varies, and then conduct correlation tests to analyze the correlation between the mutation ratios and mutation scores. Given that multiple test suites are employed for each model, we merge the mutation score data across all test suites to obtain comprehensive results for all test suites. We then analyze the mutation score results separately for each mutation operator: the \mcd and \mcb operators, as introduced in Section~\ref{subsec:mutation_op}. 

\begin{figure}[ht]
    \centering
    \includegraphics[width=\linewidth]{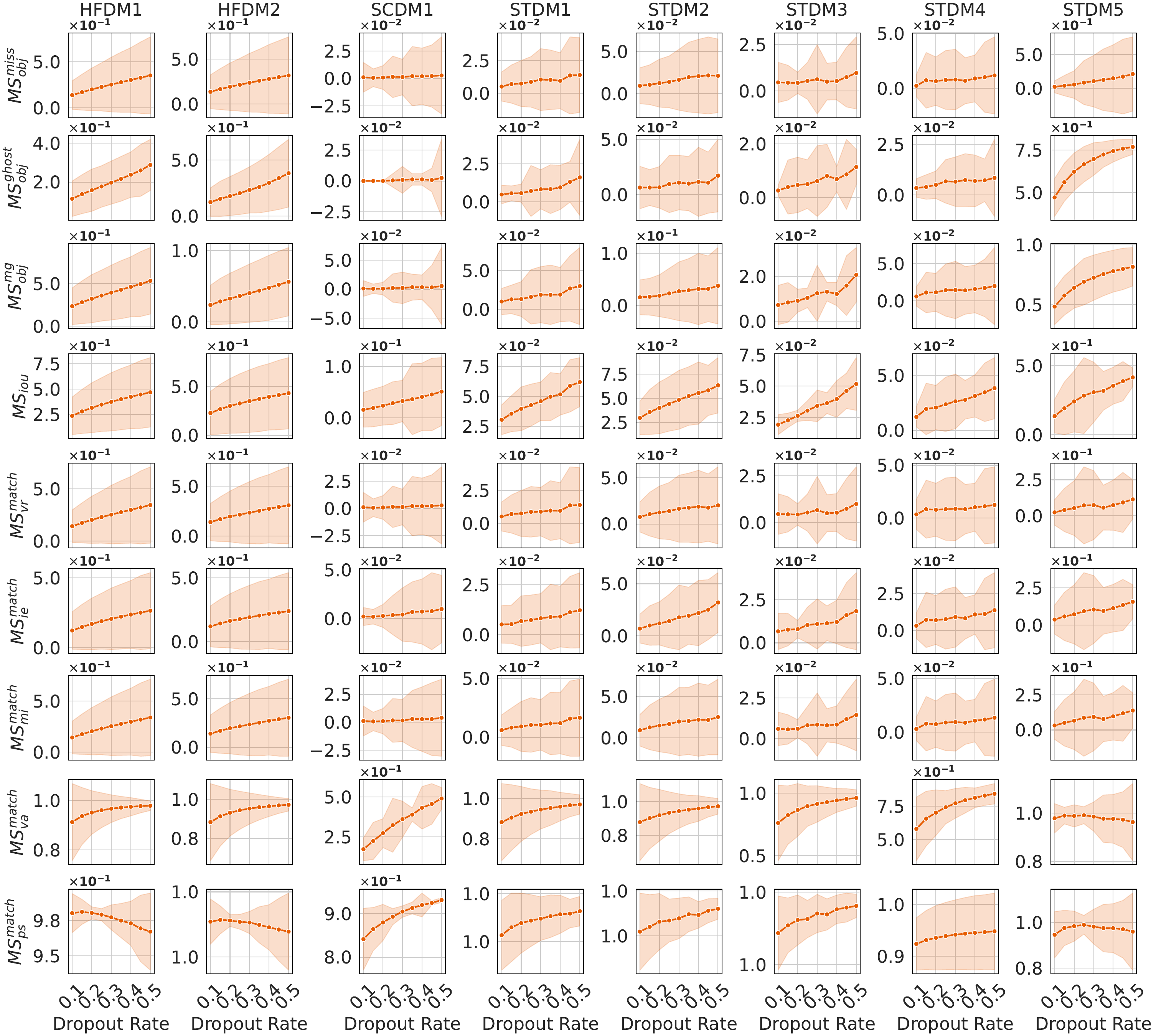}
    \caption{Mutation Scores Achieved by Test Suites with 95\% Confidence Intervals for Different Dropout Rates (Mutation Operator: \mcd, Mutation Scores: \objms, \ioums, and \uams derived from $\mathcal{S}_{match}$) -- RQ3.}
    \label{fig:mcd_obj_iou_match}
\end{figure}
\begin{figure}[ht]
    \centering
    \includegraphics[width=\linewidth]{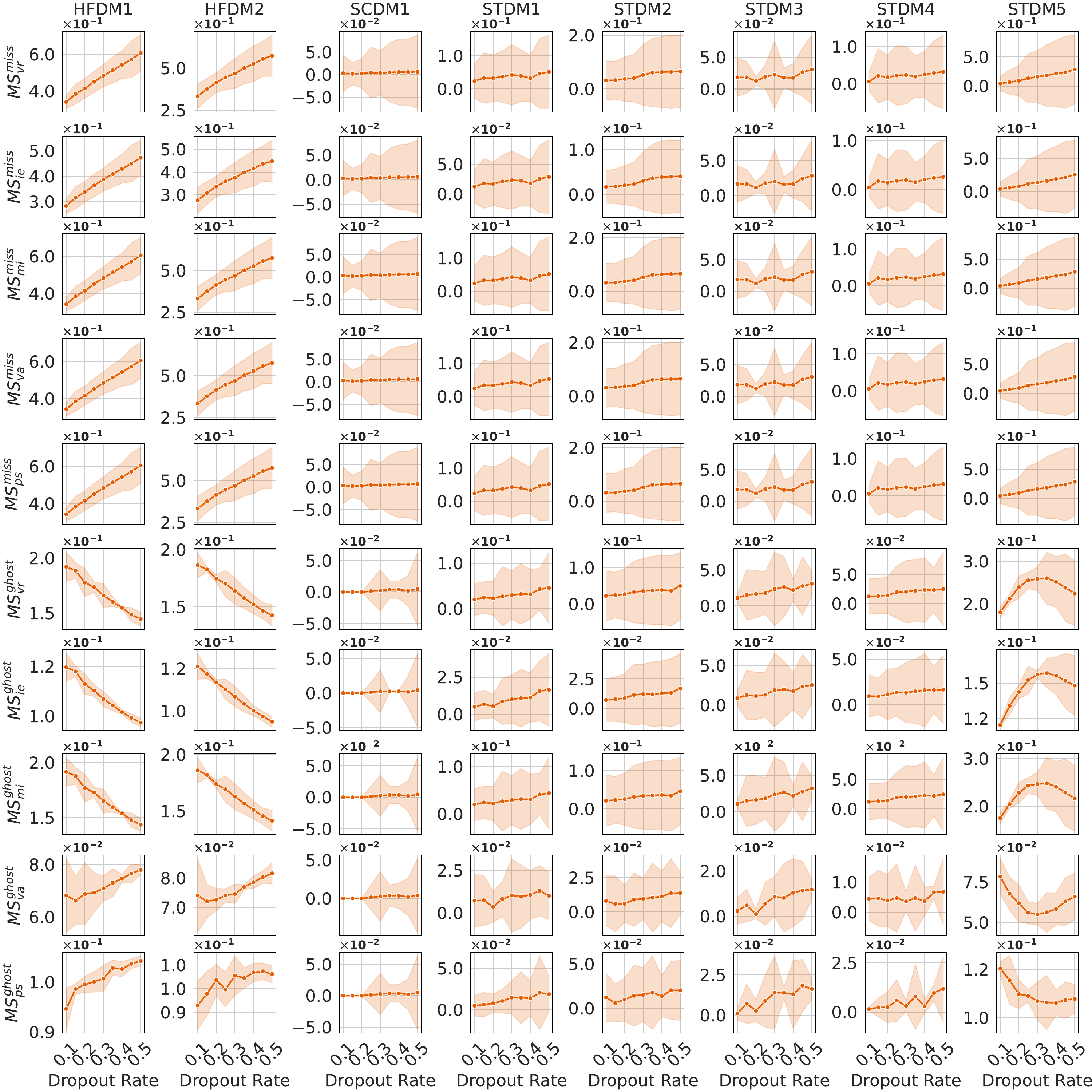}
    \caption{Mutation Scores Achieved by Test Suites with 95\% Confidence Intervals for Different Dropout Rates (Mutation Operator: \mcd, Mutation Scores: \uams Derived from $\mathcal{S}_{miss}$ and $\mathcal{S}_{ghost}$) -- RQ3.}
    \label{fig:mcd_uq_miss_ghost}
    \vspace{-4pt}
\end{figure}

\subsubsection{Analysis for \mcd Operator}
Recall from Section~\ref{subsec:mutation_op} that \mcd operator generates mutants of a given model using the MC-Dropout method~\cite{gal2016dropout}, and it has one mutation ratio called dropout rate to control the probability of neurons being dropped during inference time. A higher dropout rate indicates greater uncertainty injected into the model. Figs.~\ref{fig:mcd_obj_iou_match} and~\ref{fig:mcd_uq_miss_ghost} present the results for \mcd operator, showing the mean \gls{ms} values and their 95\% \gls{ci} across different dropout rates for each model. Table~\ref{tab:cor_test_mcd} further reports the correlation test results between dropout rate and all mutation scores using Spearman's rank correlation test. 

\textbf{\textit{Descriptive Statistics}}. As Figs.~\ref{fig:mcd_obj_iou_match} and~\ref{fig:mcd_uq_miss_ghost} show, in general, as the dropout rate increases, the values increase across the majority of mutation score metrics and models. This pattern aligns with the expectation that a higher dropout rate induces greater uncertainty, thereby inducing a higher frequency of failures in the generated mutants. Consequently, the increase in mutation scores indicates that most of them effectively capture uncertainty-induced degradation in model behavior. Furthermore, the 95\% \gls{ci} regions generally grow as the dropout rate increases, indicating higher variability in mutation score values. This suggests that while the mutation scores are sensitive to changes in model uncertainty, higher levels of uncertainty may reduce their stability.

When looking at each metric, different patterns are observed for specific mutation score metrics, which is interesting for further analysis. First, all \objms and \ioums metrics demonstrate a consistent increasing trend across all models as the dropout rate increases. This confirms their general effectiveness in detecting model performance degradation caused by the increase in model uncertainty. However, this sensitivity comes at the cost of stability, as widening \glspl{ci} is observed at higher dropout rates. 
A similar pattern is observed for \uamsmatchvr, \uamsmatchie, \uamsmatchmi, and all \uams metrics derived from $\mathcal{S}_{miss}$. Their values consistently increase across all models as the dropout rate increases, while exhibiting higher variances. This indicates that these \uams metrics are capable of capturing uncertainty-induced model performance degradation.

However, notable exceptions are observed for \uamsmatchva, \uamsmatchps, and all \uams derived from $\mathcal{S}_{ghost}$. Although \uamsmatchva and \uamsmatchps increase in certain cases (e.g., \stdm{1} and \stdm{1} to \stdm{4}), they exhibit a flat or even negative trend for models such as \hfdm{1} and \hfdm{2}. Interestingly, their \glspl{ci} become narrower as the dropout rate increases, indicating improved stability as the uncertainty increases. Given that \uamsmatchva and \uamsmatchps quantify uncertainty-aware mutation scores in regression outputs (i.e., bounding box coordinates), these results indicate that these two metrics are not effective in capturing the model quality degradation. By inspecting the detection results, we notice a reduction in the size of the match set as the dropout rate increases. A plausible explanation is that increased uncertainty causes the model to entirely miss objects that are hard to detect (shifting them to the miss set); consequently, the objects remaining in the match set are easy targets where the model maintains confident regression predictions, leading to low or stable values in \uamsmatchva and \uamsmatchps even as overall model performance degrades. 
As for \uams derived from $\mathcal{S}_{ghost}$, they exhibit different patterns for different models. Specifically, \uamsghostvr, \uamsghostie, and \uamsghostmi show the expected positive increase trend for \scdm{1} and \stdm{1} to \stdm{4}. However, for \hfdm{1}, \hfdm{2}, and \stdm{5}, they either exhibit negative correlations or no clear trend. This model-specific pattern is likely attributed to the number and characteristics of ghost objects generated by each model.

\textbf{\textit{Correlation Analysis}}. As Table~\ref{tab:cor_test_mcd} shows, \objms and \ioums metrics exhibit a statistically significant and very high positive correlation ($\rho \approx 1.000$) with the dropout rate across nearly all models. This statistically confirms that increasing the dropout rate (corresponding to greater uncertainty), consistently leads to higher mutation score values. This observation confirms that these metrics are effective in capturing the model degradation under varying uncertainty levels. 
\begin{table}[ht]
    \centering
    \caption{\textbf{Correlation Test Results between Dropout Rates and Mutation Scores for MC-Dropout-based Mutation Operator Using Spearman's Rank Correlation ($\rho$) test -- RQ3.} Bold values indicate statistically significant correlations ($p < 0.01$). The $\rho$ values are further interpreted as: \colorbox{corrVeryHigh}{Very High} if $\rho$ $\in$ [0.900, 1.000] or [-1.000, -0.900], \colorbox{corrHigh}{High} if $\rho$ $\in$ [0.700, 0.900) or (-0.900, -0.700], \colorbox{corrModerate}{Moderate} if $\rho$ $\in$ [0.50, 0.700) or (-0.700, 0.500], and \colorbox{corrLow}{Low} if $\rho$ $\in$ [0.300, 0.500) or (-0.500, -0.300].}
    \resizebox{\textwidth}{!}{% \begin{tabular}{lrrrrrrrrrrrrrrrrrrrrrrrrrrrrrrrr}

\begin{tabular}{llllllllllllllllll}
\toprule
\multicolumn{2}{c}{\multirow{2}{*}{\textit{Metrics}}} & \multicolumn{2}{c}{\hfdm{1}} & \multicolumn{2}{c}{\hfdm{2}} & \multicolumn{2}{c}{\scdm{1}} & \multicolumn{2}{c}{\stdm{1}} & \multicolumn{2}{c}{\stdm{2}} & \multicolumn{2}{c}{\stdm{3}} & \multicolumn{2}{c}{\stdm{4}} & \multicolumn{2}{c}{\stdm{5}} \\
 & & $p$ & $\rho$ & $p$ & $\rho$ & $p$ & $\rho$ & $p$ & $\rho$ & $p$ & $\rho$ & $p$ & $\rho$ & $p$ & $\rho$ & $p$ & $\rho$ \\
\midrule
\multirow{3}{*}{\textit{Obj-MS}} & $MS_{obj}^{miss}$ & \textbf{<0.01} & \cellcolor{corrVeryHigh}\textbf{1.000} & \textbf{<0.01} & \cellcolor{corrVeryHigh}\textbf{1.000} & \textbf{<0.01} & \cellcolor{corrVeryHigh}\textbf{0.917} & \textbf{<0.01} & \cellcolor{corrVeryHigh}\textbf{0.933} & \textbf{<0.01} & \cellcolor{corrVeryHigh}\textbf{0.983} & \textbf{<0.01} & \cellcolor{corrHigh}\textbf{0.800} & \textbf{<0.01} & \cellcolor{corrHigh}\textbf{0.867} & \textbf{<0.01} & \cellcolor{corrVeryHigh}\textbf{1.000} \\
 & $MS_{obj}^{ghost}$ & \textbf{<0.01} & \cellcolor{corrVeryHigh}\textbf{1.000} & \textbf{<0.01} & \cellcolor{corrVeryHigh}\textbf{1.000} & \textbf{<0.01} & \cellcolor{corrHigh}\textbf{0.877} & \textbf{<0.01} & \cellcolor{corrVeryHigh}\textbf{1.000} & \textbf{<0.01} & \cellcolor{corrVeryHigh}\textbf{0.917} & \textbf{<0.01} & \cellcolor{corrVeryHigh}\textbf{0.983} & \textbf{<0.01} & \cellcolor{corrVeryHigh}\textbf{0.933} & \textbf{<0.01} & \cellcolor{corrVeryHigh}\textbf{1.000} \\
 & $MS_{obj}^{mg}$ & \textbf{<0.01} & \cellcolor{corrVeryHigh}\textbf{1.000} & \textbf{<0.01} & \cellcolor{corrVeryHigh}\textbf{1.000} & \textbf{<0.01} & \cellcolor{corrHigh}\textbf{0.883} & \textbf{<0.01} & \cellcolor{corrVeryHigh}\textbf{0.983} & \textbf{<0.01} & \cellcolor{corrVeryHigh}\textbf{1.000} & \textbf{<0.01} & \cellcolor{corrVeryHigh}\textbf{0.950} & \textbf{<0.01} & \cellcolor{corrVeryHigh}\textbf{0.950} & \textbf{<0.01} & \cellcolor{corrVeryHigh}\textbf{1.000} \\ \midrule
\textit{IoU-MS} & $MS_{iou}$ & \textbf{<0.01} & \cellcolor{corrVeryHigh}\textbf{1.000} & \textbf{<0.01} & \cellcolor{corrVeryHigh}\textbf{1.000} & \textbf{<0.01} & \cellcolor{corrVeryHigh}\textbf{1.000} & \textbf{<0.01} & \cellcolor{corrVeryHigh}\textbf{1.000} & \textbf{<0.01} & \cellcolor{corrVeryHigh}\textbf{1.000} & \textbf{<0.01} & \cellcolor{corrVeryHigh}\textbf{1.000} & \textbf{<0.01} & \cellcolor{corrVeryHigh}\textbf{1.000} & \textbf{<0.01} & \cellcolor{corrVeryHigh}\textbf{1.000} \\ \midrule
\multirow{15}{*}{\textit{UA-MS}} & $MS_{vr}^{match}$ & \textbf{<0.01} & \cellcolor{corrVeryHigh}\textbf{1.000} & \textbf{<0.01} & \cellcolor{corrVeryHigh}\textbf{1.000} & \textbf{<0.01} & \cellcolor{corrVeryHigh}\textbf{0.917} & \textbf{<0.01} & \cellcolor{corrVeryHigh}\textbf{0.983} & \textbf{<0.01} & \cellcolor{corrVeryHigh}\textbf{0.983} & \textbf{<0.01} & \cellcolor{corrHigh}\textbf{0.800} & \textbf{<0.01} & \cellcolor{corrVeryHigh}\textbf{0.933} & \textbf{<0.01} & \cellcolor{corrVeryHigh}\textbf{0.917} \\
 & $MS_{ie}^{match}$ & \textbf{<0.01} & \cellcolor{corrVeryHigh}\textbf{1.000} & \textbf{<0.01} & \cellcolor{corrVeryHigh}\textbf{1.000} & \textbf{<0.01} & \cellcolor{corrVeryHigh}\textbf{0.983} & \textbf{<0.01} & \cellcolor{corrVeryHigh}\textbf{1.000} & \textbf{<0.01} & \cellcolor{corrVeryHigh}\textbf{1.000} & \textbf{<0.01} & \cellcolor{corrVeryHigh}\textbf{1.000} & \textbf{<0.01} & \cellcolor{corrVeryHigh}\textbf{0.967} & \textbf{<0.01} & \cellcolor{corrVeryHigh}\textbf{0.983} \\
 & $MS_{mi}^{match}$ & \textbf{<0.01} & \cellcolor{corrVeryHigh}\textbf{1.000} & \textbf{<0.01} & \cellcolor{corrVeryHigh}\textbf{1.000} & \textbf{<0.01} & \cellcolor{corrHigh}\textbf{0.883} & \textbf{<0.01} & \cellcolor{corrVeryHigh}\textbf{1.000} & \textbf{<0.01} & \cellcolor{corrVeryHigh}\textbf{0.983} & \textbf{<0.01} & \cellcolor{corrVeryHigh}\textbf{0.933} & \textbf{<0.01} & \cellcolor{corrVeryHigh}\textbf{0.933} & \textbf{<0.01} & \cellcolor{corrVeryHigh}\textbf{0.950} \\
 & $MS_{va}^{match}$ & \textbf{<0.01} & \cellcolor{corrVeryHigh}\textbf{1.000} & \textbf{<0.01} & \cellcolor{corrVeryHigh}\textbf{1.000} & \textbf{<0.01} & \cellcolor{corrVeryHigh}\textbf{1.000} & \textbf{<0.01} & \cellcolor{corrVeryHigh}\textbf{1.000} & \textbf{<0.01} & \cellcolor{corrVeryHigh}\textbf{1.000} & \textbf{<0.01} & \cellcolor{corrVeryHigh}\textbf{1.000} & \textbf{<0.01} & \cellcolor{corrVeryHigh}\textbf{1.000} & 0.013 & -0.783 \\
 & $MS_{ps}^{match}$ & \textbf{<0.01} & \cellcolor{corrVeryHigh}\textbf{-0.950} & \textbf{<0.01} & \cellcolor{corrVeryHigh}\textbf{-0.950} & \textbf{<0.01} & \cellcolor{corrVeryHigh}\textbf{1.000} & \textbf{<0.01} & \cellcolor{corrVeryHigh}\textbf{1.000} & \textbf{<0.01} & \cellcolor{corrVeryHigh}\textbf{0.983} & \textbf{<0.01} & \cellcolor{corrVeryHigh}\textbf{0.983} & \textbf{<0.01} & \cellcolor{corrVeryHigh}\textbf{1.000} & 0.460 & -0.283 \\ \cmidrule(lr){2-18}
 & $MS_{vr}^{miss}$ & \textbf{<0.01} & \cellcolor{corrVeryHigh}\textbf{1.000} & \textbf{<0.01} & \cellcolor{corrVeryHigh}\textbf{1.000} & \textbf{<0.01} & \cellcolor{corrVeryHigh}\textbf{0.929} & 0.030 & 0.717 & \textbf{<0.01} & \cellcolor{corrVeryHigh}\textbf{1.000} & 0.224 & 0.450 & \textbf{<0.01} & \cellcolor{corrHigh}\textbf{0.867} & \textbf{<0.01} & \cellcolor{corrVeryHigh}\textbf{1.000} \\
 & $MS_{ie}^{miss}$ & \textbf{<0.01} & \cellcolor{corrVeryHigh}\textbf{1.000} & \textbf{<0.01} & \cellcolor{corrVeryHigh}\textbf{1.000} & \textbf{<0.01} & \cellcolor{corrVeryHigh}\textbf{0.933} & 0.013 & 0.783 & \textbf{<0.01} & \cellcolor{corrVeryHigh}\textbf{1.000} & 0.205 & 0.467 & \textbf{<0.01} & \cellcolor{corrHigh}\textbf{0.867} & \textbf{<0.01} & \cellcolor{corrVeryHigh}\textbf{1.000} \\
 & $MS_{mi}^{miss}$ & \textbf{<0.01} & \cellcolor{corrVeryHigh}\textbf{1.000} & \textbf{<0.01} & \cellcolor{corrVeryHigh}\textbf{1.000} & \textbf{<0.01} & \cellcolor{corrVeryHigh}\textbf{0.933} & 0.030 & 0.717 & \textbf{<0.01} & \cellcolor{corrVeryHigh}\textbf{1.000} & 0.224 & 0.450 & \textbf{<0.01} & \cellcolor{corrHigh}\textbf{0.867} & \textbf{<0.01} & \cellcolor{corrVeryHigh}\textbf{1.000} \\
 & $MS_{va}^{miss}$ & \textbf{<0.01} & \cellcolor{corrVeryHigh}\textbf{1.000} & \textbf{<0.01} & \cellcolor{corrVeryHigh}\textbf{1.000} & \textbf{<0.01} & \cellcolor{corrVeryHigh}\textbf{0.929} & 0.030 & 0.717 & \textbf{<0.01} & \cellcolor{corrVeryHigh}\textbf{1.000} & 0.224 & 0.450 & \textbf{<0.01} & \cellcolor{corrHigh}\textbf{0.867} & \textbf{<0.01} & \cellcolor{corrVeryHigh}\textbf{1.000} \\
 & $MS_{ps}^{miss}$ & \textbf{<0.01} & \cellcolor{corrVeryHigh}\textbf{1.000} & \textbf{<0.01} & \cellcolor{corrVeryHigh}\textbf{1.000} & \textbf{<0.01} & \cellcolor{corrVeryHigh}\textbf{0.929} & 0.030 & 0.717 & \textbf{<0.01} & \cellcolor{corrVeryHigh}\textbf{1.000} & 0.224 & 0.450 & \textbf{<0.01} & \cellcolor{corrHigh}\textbf{0.867} & \textbf{<0.01} & \cellcolor{corrVeryHigh}\textbf{1.000} \\ \cmidrule(lr){2-18}
 & $MS_{vr}^{ghost}$ & \textbf{<0.01} & \cellcolor{corrVeryHigh}\textbf{-1.000} & \textbf{<0.01} & \cellcolor{corrVeryHigh}\textbf{-1.000} & \textbf{<0.01} & \cellcolor{corrHigh}\textbf{0.877} & \textbf{<0.01} & \cellcolor{corrVeryHigh}\textbf{0.967} & \textbf{<0.01} & \cellcolor{corrVeryHigh}\textbf{0.950} & \textbf{<0.01} & \cellcolor{corrVeryHigh}\textbf{0.950} & \textbf{<0.01} & \cellcolor{corrVeryHigh}\textbf{0.983} & 0.433 & 0.300 \\
 & $MS_{ie}^{ghost}$ & \textbf{<0.01} & \cellcolor{corrVeryHigh}\textbf{-1.000} & \textbf{<0.01} & \cellcolor{corrVeryHigh}\textbf{-1.000} & \textbf{<0.01} & \cellcolor{corrHigh}\textbf{0.865} & \textbf{<0.01} & \cellcolor{corrVeryHigh}\textbf{0.983} & \textbf{<0.01} & \cellcolor{corrVeryHigh}\textbf{0.983} & \textbf{<0.01} & \cellcolor{corrVeryHigh}\textbf{0.933} & \textbf{<0.01} & \cellcolor{corrVeryHigh}\textbf{0.967} & 0.139 & 0.533 \\
 & $MS_{mi}^{ghost}$ & \textbf{<0.01} & \cellcolor{corrVeryHigh}\textbf{-1.000} & \textbf{<0.01} & \cellcolor{corrVeryHigh}\textbf{-1.000} & \textbf{<0.01} & \cellcolor{corrHigh}\textbf{0.865} & \textbf{<0.01} & \cellcolor{corrVeryHigh}\textbf{0.967} & \textbf{<0.01} & \cellcolor{corrVeryHigh}\textbf{0.950} & \textbf{<0.01} & \cellcolor{corrVeryHigh}\textbf{0.950} & \textbf{<0.01} & \cellcolor{corrVeryHigh}\textbf{0.983} & 0.308 & 0.383 \\
 & $MS_{va}^{ghost}$ & \textbf{<0.01} & \cellcolor{corrVeryHigh}\textbf{0.983} & \textbf{<0.01} & \cellcolor{corrVeryHigh}\textbf{0.950} & \textbf{<0.01} & \cellcolor{corrHigh}\textbf{0.865} & \textbf{<0.01} & \cellcolor{corrHigh}\textbf{0.817} & \textbf{<0.01} & \cellcolor{corrVeryHigh}\textbf{0.950} & \textbf{<0.01} & \cellcolor{corrVeryHigh}\textbf{0.933} & 0.205 & 0.467 & 0.516 & -0.250 \\
 & $MS_{ps}^{ghost}$ & \textbf{<0.01} & \cellcolor{corrVeryHigh}\textbf{0.983} & \textbf{<0.01} & \cellcolor{corrVeryHigh}\textbf{0.917} & \textbf{<0.01} & \cellcolor{corrHigh}\textbf{0.877} & \textbf{<0.01} & \cellcolor{corrVeryHigh}\textbf{0.917} & \textbf{<0.01} & \cellcolor{corrHigh}\textbf{0.833} & \textbf{<0.01} & \cellcolor{corrVeryHigh}\textbf{0.900} & \textbf{<0.01} & \cellcolor{corrHigh}\textbf{0.883} & 0.025 & -0.733 \\
\bottomrule
\end{tabular}}
    \label{tab:cor_test_mcd}
\end{table}

The correlation results for \uams reveal different patterns for specific metrics. For \uams derived from $\mathcal{S}_{match}$, \uamsmatchvr, \uamsmatchie, \uamsmatchmi, and \uamsmatchva exhibit statistically significant and very high positive correlations across nearly all models. A notable exception is \uamsmatchps for \hfdm{1} and \hfdm{2}, which shows a strong negative correlation ($\rho = -0.950$), and this aligns with the prior observation from Figure~\ref{fig:mcd_obj_iou_match} and~\ref{fig:mcd_uq_miss_ghost}.
For the \uams derived from $\mathcal{S}_{miss}$, statistically significant and high or very high positive correlations are observed for the majority of models (e.g., \hfdm{1}, \hfdm{2}, \scdm{1}, and \stdm{2, 4, 5}). 
This indicates that these metrics are effective in capturing the increased uncertainty-induced failures, particularly those related to missed detections.
However, \stdm{1} and \stdm{3} exhibit a deviation where the correlations are not statistically significant ($p > 0.05$), indicating limited effectiveness in capturing model failures as the increase of uncertainty for these two specific models. 
Finally, the Ghost set metrics highlight the most significant architectural disparities under varying uncertainty levels. While \scdm{1} and most \stdm{} models show positive correlations, \hfdm{1} and \hfdm{2} exhibit a perfect negative correlation ($\rho = -1.000$) for \uamsghostvr, \uamsghostie, and \uamsghostmi. 
Interestingly, this trend reverses for \uamsghostva and \uamsghostps in the same \hfdm{} models, where the correlation is strongly positive.
Recall that \uamsghostvr, \uamsghostie, and \uamsghostmi are derived from classification uncertainty, whereas \uamsghostva and \uamsghostps are derived from regression uncertainty. This indicates that as overall model uncertainty increases, the classification predictions of ghost objects remain relatively stable, while the bounding box predictions become more unstable.

\subsubsection{Analysis for \mcb Operator}
\begin{figure}[ht]
    \centering
    \includegraphics[width=\linewidth]{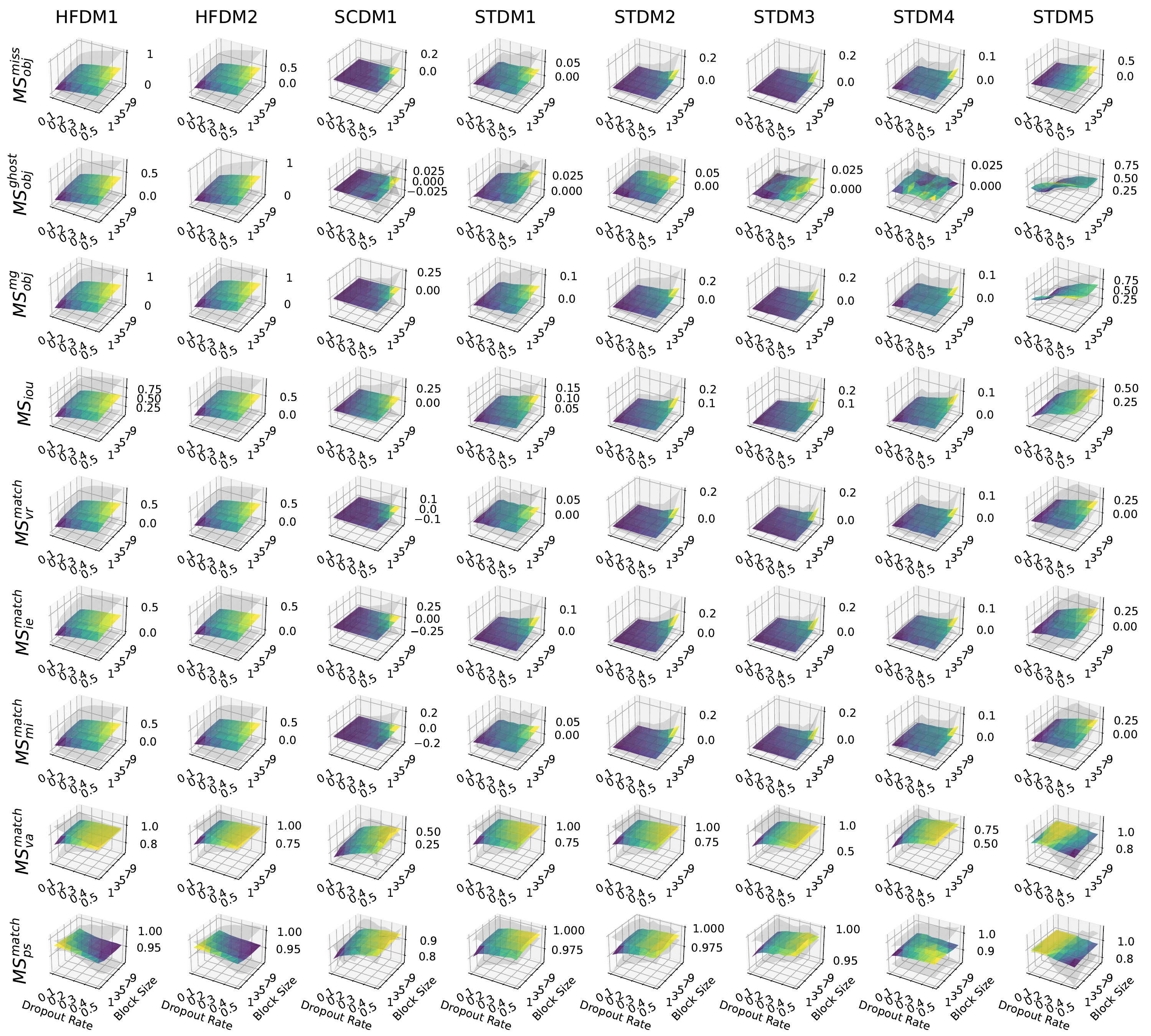}
    \caption{Mutation Scores Achieved by Test Suites with 95\% Confidence Intervals for Dropout Rates $\times$ Block Sizes (Mutation Operator: \mcb, Mutation Scores: \objms, \ioums, and \uams derived from $\mathcal{S}_{match}$) -- RQ3.}
    \label{fig:mcb_3d_obj_iou_match}
\end{figure}

\begin{figure}[ht]
    \centering
    \includegraphics[width=\linewidth]{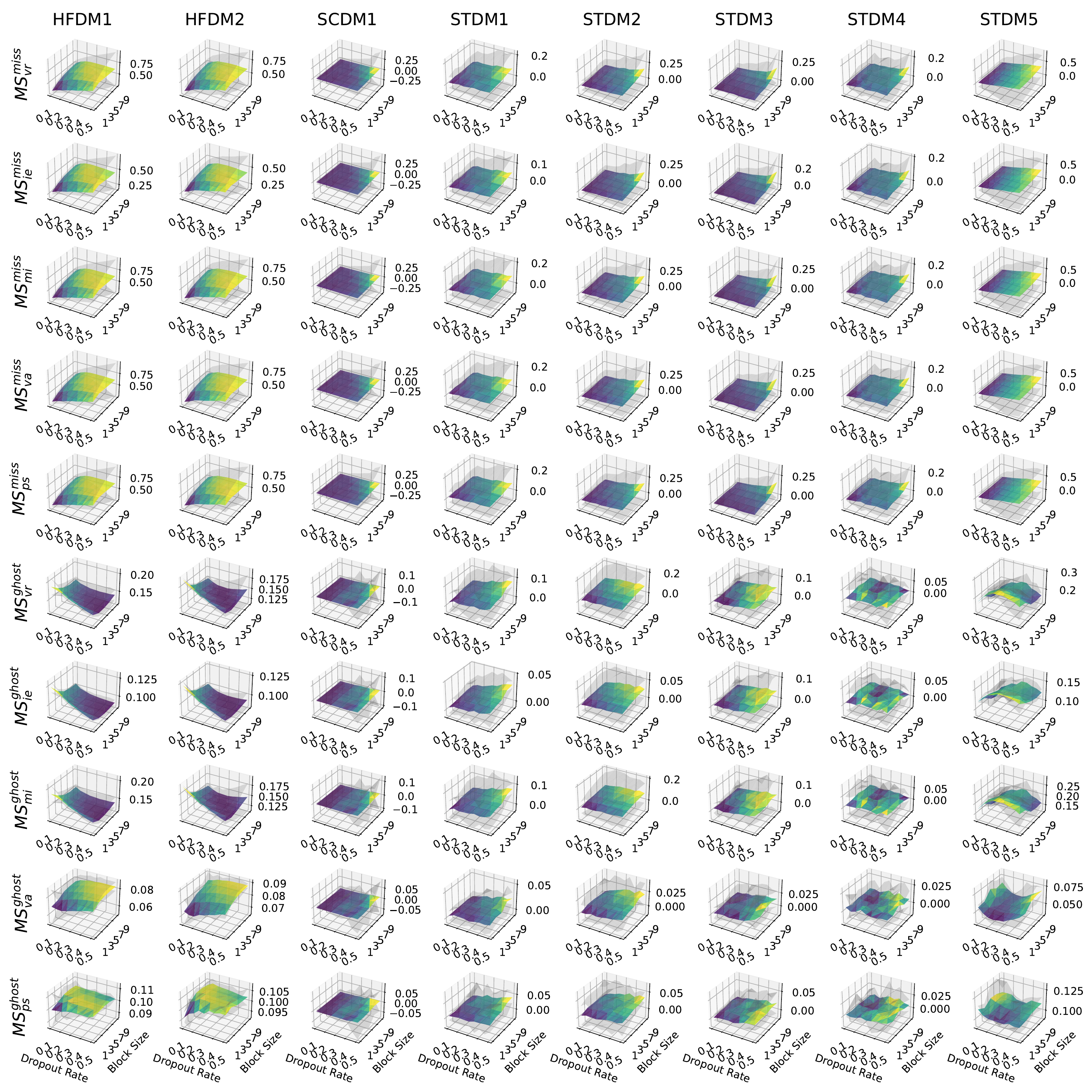}
    \caption{Mutation Scores Achieved by Test Suites with 95\% Confidence Intervals for Dropout Rates $\times$ Block Sizes (Mutation Operator: \mcb, Mutation Scores: \uams Derived from $\mathcal{S}_{miss}$ and $\mathcal{S}_{ghost}$) -- RQ3.}
    \label{fig:mcb_3d_uq_miss_ghost}
    \vspace{-6pt}
\end{figure}

\mcb operator generates mutants of a given model using the MC-DropBlock method~\cite{yelleni2024monte}. Unlike the \mcd operator, which is governed by a single mutation ratio, \mcb operator employs two mutation ratios: the dropout rate determining the proportion of activations perturbed, and the block size, controlling the spatial granularity of the injected uncertainty. Higher mutation ratios indicate a greater injected uncertainty. Figs.~\ref{fig:mcb_3d_obj_iou_match} and~\ref{fig:mcb_3d_uq_miss_ghost} present the results for the \mcb operator, illustrating how each mutation score varies across different combinations of dropout rate and block size for each model. Table~\ref{tab:cor_test_mcd} further reports the multiple correlation results between mutation scores and the combinations of dropout rate and block size. 

\textbf{\textit{Descriptive Statistics}}. As shown in Figs.~\ref{fig:mcb_3d_obj_iou_match} and~\ref{fig:mcb_3d_uq_miss_ghost}, mutation scores generally increase with both dropout rate and block size, indicating their effectiveness in capturing uncertainty-induced degradation in model performance. Besides, among the two mutation ratios, block size appears to have a greater impact than dropout rate, often leading to steeper gradients in mutation score surfaces. 
At the level of individual metric, both \objms and \ioums metrics exhibit clear monotonic gradients, with surfaces sloping upward toward regions corresponding to larger block sizes and higher dropout rates. This behavior highlights the effectiveness and sensitivity of \objms and \ioums metrics in capturing uncertainty-induced failures under varying uncertainty levels. 

Similarly, \uams demonstrates consistent upward trends across \uamsmatchvr, \uamsmatchie, and \uamsmatchmi, \uamsmatchva, as well as across all \uams metrics derived from $\mathcal{S}_{miss}$, confirming that these metrics reliably capture model uncertainty. However, for \uamsmatchps, and all \uams derived from $\mathcal{S}_{ghost}$, the surfaces exhibit flatter patterns and no distinct monotonic gradients are observed. This behavior suggests that these metrics are less effective in capturing the model's performance degradation as uncertainty increases. A possible reason might relate to the detection nature of the ghost and miss objects, that ghost objects act as false positives driven by random noise; their internal uncertainty estimates tend to be irregular and do not scale linearly with the mutation ratios.
\begin{table}[ht]
    \centering
    \caption{\textbf{Multiple Correlation Test Results between (Dropout Rate, Block Size) and Mutation Scores for \mcb Operator Using Multiple Correlation Coefficient ($R$) -- RQ3.} Bold values indicate statistically significant correlations ($p < 0.01$). The $R$ values are further interpreted as: \colorbox{corrVeryHigh}{Very High} if $R$ $\in$ [0.900, 1.000], \colorbox{corrHigh}{High} if $R$ $\in$ [0.700, 0.900), \colorbox{corrModerate}{Moderate} if $R$ $\in$ [0.50, 0.700), and \colorbox{corrLow}{Low} if $R$ $\in$ [0.300, 0.500).}
    \resizebox{\textwidth}{!}{\begin{tabular}{llllllllllllllllll}
\toprule
\multicolumn{2}{c}{\multirow{2}{*}{\textit{Metrics}}} & \multicolumn{2}{c}{\hfdm{1}} & \multicolumn{2}{c}{\hfdm{2}} & \multicolumn{2}{c}{\scdm{1}} & \multicolumn{2}{c}{\stdm{1}} & \multicolumn{2}{c}{\stdm{2}} & \multicolumn{2}{c}{\stdm{3}} & \multicolumn{2}{c}{\stdm{4}} & \multicolumn{2}{c}{\stdm{5}} \\
& & $p$ & $R$ & $p$ & $R$ & $p$ & $R$ & $p$ & $R$ & $p$ & $R$ & $p$ & $R$ & $p$ & $R$ & $p$ & $R$ \\
\midrule
\multirow{3}{*}{\textit{Obj-MS}} & $MS_{obj}^{miss}$ & \textbf{<0.01} & \cellcolor{corrVeryHigh}\textbf{0.974} & \textbf{<0.01} & \cellcolor{corrVeryHigh}\textbf{0.976} & \textbf{<0.01} & \cellcolor{corrHigh}\textbf{0.814} & \textbf{<0.01} & \cellcolor{corrVeryHigh}\textbf{0.951} & \textbf{<0.01} & \cellcolor{corrHigh}\textbf{0.805} & \textbf{<0.01} & \cellcolor{corrHigh}\textbf{0.751} & \textbf{<0.01} & \cellcolor{corrHigh}\textbf{0.829} & \textbf{<0.01} & \cellcolor{corrVeryHigh}\textbf{0.984} \\
& $MS_{obj}^{ghost}$ & \textbf{<0.01} & \cellcolor{corrVeryHigh}\textbf{0.976} & \textbf{<0.01} & \cellcolor{corrVeryHigh}\textbf{0.985} & \textbf{<0.01} & \cellcolor{corrHigh}\textbf{0.856} & \textbf{<0.01} & \cellcolor{corrVeryHigh}\textbf{0.952} & \textbf{<0.01} & \cellcolor{corrVeryHigh}\textbf{0.966} & \textbf{<0.01} & \cellcolor{corrHigh}\textbf{0.898} & 0.013 & 0.432 & \textbf{<0.01} & \cellcolor{corrVeryHigh}\textbf{0.917} \\
& $MS_{obj}^{mg}$ & \textbf{<0.01} & \cellcolor{corrVeryHigh}\textbf{0.969} & \textbf{<0.01} & \cellcolor{corrVeryHigh}\textbf{0.979} & \textbf{<0.01} & \cellcolor{corrHigh}\textbf{0.846} & \textbf{<0.01} & \cellcolor{corrVeryHigh}\textbf{0.966} & \textbf{<0.01} & \cellcolor{corrVeryHigh}\textbf{0.903} & \textbf{<0.01} & \cellcolor{corrHigh}\textbf{0.800} & \textbf{<0.01} & \cellcolor{corrHigh}\textbf{0.849} & \textbf{<0.01} & \cellcolor{corrVeryHigh}\textbf{0.924} \\ \midrule
\textit{IoU-MS} & $MS_{iou}$ & \textbf{<0.01} & \cellcolor{corrVeryHigh}\textbf{0.957} & \textbf{<0.01} & \cellcolor{corrVeryHigh}\textbf{0.968} & \textbf{<0.01} & \cellcolor{corrVeryHigh}\textbf{0.987} & \textbf{<0.01} & \cellcolor{corrVeryHigh}\textbf{0.977} & \textbf{<0.01} & \cellcolor{corrVeryHigh}\textbf{0.915} & \textbf{<0.01} & \cellcolor{corrHigh}\textbf{0.898} & \textbf{<0.01} & \cellcolor{corrVeryHigh}\textbf{0.951} & \textbf{<0.01} & \cellcolor{corrVeryHigh}\textbf{0.977} \\ \midrule
\multirow{15}{*}{\textit{UA-MS}} & $MS_{vr}^{match}$ & \textbf{<0.01} & \cellcolor{corrVeryHigh}\textbf{0.970} & \textbf{<0.01} & \cellcolor{corrVeryHigh}\textbf{0.975} & \textbf{<0.01} & \cellcolor{corrHigh}\textbf{0.792} & \textbf{<0.01} & \cellcolor{corrVeryHigh}\textbf{0.930} & \textbf{<0.01} & \cellcolor{corrHigh}\textbf{0.764} & \textbf{<0.01} & \cellcolor{corrHigh}\textbf{0.736} & \textbf{<0.01} & \cellcolor{corrHigh}\textbf{0.834} & \textbf{<0.01} & \cellcolor{corrVeryHigh}\textbf{0.947} \\
& $MS_{ie}^{match}$ & \textbf{<0.01} & \cellcolor{corrVeryHigh}\textbf{0.947} & \textbf{<0.01} & \cellcolor{corrVeryHigh}\textbf{0.960} & \textbf{<0.01} & \cellcolor{corrHigh}\textbf{0.883} & \textbf{<0.01} & \cellcolor{corrHigh}\textbf{0.897} & \textbf{<0.01} & \cellcolor{corrHigh}\textbf{0.837} & \textbf{<0.01} & \cellcolor{corrHigh}\textbf{0.851} & \textbf{<0.01} & \cellcolor{corrVeryHigh}\textbf{0.909} & \textbf{<0.01} & \cellcolor{corrVeryHigh}\textbf{0.967} \\
& $MS_{mi}^{match}$ & \textbf{<0.01} & \cellcolor{corrVeryHigh}\textbf{0.967} & \textbf{<0.01} & \cellcolor{corrVeryHigh}\textbf{0.973} & \textbf{<0.01} & \cellcolor{corrHigh}\textbf{0.827} & \textbf{<0.01} & \cellcolor{corrVeryHigh}\textbf{0.949} & \textbf{<0.01} & \cellcolor{corrHigh}\textbf{0.802} & \textbf{<0.01} & \cellcolor{corrHigh}\textbf{0.778} & \textbf{<0.01} & \cellcolor{corrHigh}\textbf{0.862} & \textbf{<0.01} & \cellcolor{corrVeryHigh}\textbf{0.958} \\
& $MS_{va}^{match}$ & \textbf{<0.01} & \cellcolor{corrHigh}\textbf{0.810} & \textbf{<0.01} & \cellcolor{corrHigh}\textbf{0.882} & \textbf{<0.01} & \cellcolor{corrVeryHigh}\textbf{0.961} & \textbf{<0.01} & \cellcolor{corrVeryHigh}\textbf{0.917} & \textbf{<0.01} & \cellcolor{corrVeryHigh}\textbf{0.923} & \textbf{<0.01} & \cellcolor{corrVeryHigh}\textbf{0.927} & \textbf{<0.01} & \cellcolor{corrVeryHigh}\textbf{0.902} & \textbf{<0.01} & \cellcolor{corrHigh}\textbf{0.767} \\
& $MS_{ps}^{match}$ & \textbf{<0.01} & \cellcolor{corrVeryHigh}\textbf{0.935} & \textbf{<0.01} & \cellcolor{corrVeryHigh}\textbf{0.940} & \textbf{<0.01} & \cellcolor{corrVeryHigh}\textbf{0.925} & \textbf{<0.01} & \cellcolor{corrVeryHigh}\textbf{0.925} & \textbf{<0.01} & \cellcolor{corrVeryHigh}\textbf{0.910} & \textbf{<0.01} & \cellcolor{corrVeryHigh}\textbf{0.913} & \textbf{<0.01} & \cellcolor{corrModerate}\textbf{0.518} & \textbf{<0.01} & \cellcolor{corrVeryHigh}\textbf{0.916} \\ \cmidrule(lr){2-18}
& $MS_{vr}^{miss}$ & \textbf{<0.01} & \cellcolor{corrHigh}\textbf{0.861} & \textbf{<0.01} & \cellcolor{corrHigh}\textbf{0.860} & \textbf{<0.01} & \cellcolor{corrHigh}\textbf{0.831} & \textbf{<0.01} & \cellcolor{corrVeryHigh}\textbf{0.927} & \textbf{<0.01} & \cellcolor{corrVeryHigh}\textbf{0.909} & \textbf{<0.01} & \cellcolor{corrHigh}\textbf{0.825} & \textbf{<0.01} & \cellcolor{corrHigh}\textbf{0.848} & \textbf{<0.01} & \cellcolor{corrVeryHigh}\textbf{0.990} \\
& $MS_{ie}^{miss}$ & \textbf{<0.01} & \cellcolor{corrHigh}\textbf{0.798} & \textbf{<0.01} & \cellcolor{corrHigh}\textbf{0.786} & \textbf{<0.01} & \cellcolor{corrHigh}\textbf{0.826} & \textbf{<0.01} & \cellcolor{corrVeryHigh}\textbf{0.929} & \textbf{<0.01} & \cellcolor{corrHigh}\textbf{0.869} & \textbf{<0.01} & \cellcolor{corrHigh}\textbf{0.819} & \textbf{<0.01} & \cellcolor{corrHigh}\textbf{0.843} & \textbf{<0.01} & \cellcolor{corrVeryHigh}\textbf{0.990} \\
& $MS_{mi}^{miss}$ & \textbf{<0.01} & \cellcolor{corrHigh}\textbf{0.861} & \textbf{<0.01} & \cellcolor{corrHigh}\textbf{0.860} & \textbf{<0.01} & \cellcolor{corrHigh}\textbf{0.831} & \textbf{<0.01} & \cellcolor{corrVeryHigh}\textbf{0.927} & \textbf{<0.01} & \cellcolor{corrVeryHigh}\textbf{0.909} & \textbf{<0.01} & \cellcolor{corrHigh}\textbf{0.825} & \textbf{<0.01} & \cellcolor{corrHigh}\textbf{0.848} & \textbf{<0.01} & \cellcolor{corrVeryHigh}\textbf{0.990} \\
& $MS_{va}^{miss}$ & \textbf{<0.01} & \cellcolor{corrHigh}\textbf{0.861} & \textbf{<0.01} & \cellcolor{corrHigh}\textbf{0.860} & \textbf{<0.01} & \cellcolor{corrHigh}\textbf{0.831} & \textbf{<0.01} & \cellcolor{corrVeryHigh}\textbf{0.927} & \textbf{<0.01} & \cellcolor{corrVeryHigh}\textbf{0.909} & \textbf{<0.01} & \cellcolor{corrHigh}\textbf{0.825} & \textbf{<0.01} & \cellcolor{corrHigh}\textbf{0.848} & \textbf{<0.01} & \cellcolor{corrVeryHigh}\textbf{0.990} \\
& $MS_{ps}^{miss}$ & \textbf{<0.01} & \cellcolor{corrHigh}\textbf{0.861} & \textbf{<0.01} & \cellcolor{corrHigh}\textbf{0.860} & \textbf{<0.01} & \cellcolor{corrHigh}\textbf{0.831} & \textbf{<0.01} & \cellcolor{corrVeryHigh}\textbf{0.927} & \textbf{<0.01} & \cellcolor{corrVeryHigh}\textbf{0.909} & \textbf{<0.01} & \cellcolor{corrHigh}\textbf{0.825} & \textbf{<0.01} & \cellcolor{corrHigh}\textbf{0.848} & \textbf{<0.01} & \cellcolor{corrVeryHigh}\textbf{0.990} \\ \cmidrule(lr){2-18}
& $MS_{vr}^{ghost}$ & \textbf{<0.01} & \cellcolor{corrHigh}\textbf{0.831} & \textbf{<0.01} & \cellcolor{corrHigh}\textbf{0.743} & \textbf{<0.01} & \cellcolor{corrHigh}\textbf{0.848} & \textbf{<0.01} & \cellcolor{corrVeryHigh}\textbf{0.932} & \textbf{<0.01} & \cellcolor{corrVeryHigh}\textbf{0.963} & \textbf{<0.01} & \cellcolor{corrHigh}\textbf{0.887} & 0.037 & 0.381 & \textbf{<0.01} & \cellcolor{corrModerate}\textbf{0.656} \\
& $MS_{ie}^{ghost}$ & \textbf{<0.01} & \cellcolor{corrHigh}\textbf{0.881} & \textbf{<0.01} & \cellcolor{corrHigh}\textbf{0.849} & \textbf{<0.01} & \cellcolor{corrHigh}\textbf{0.863} & \textbf{<0.01} & \cellcolor{corrVeryHigh}\textbf{0.950} & \textbf{<0.01} & \cellcolor{corrVeryHigh}\textbf{0.962} & \textbf{<0.01} & \cellcolor{corrHigh}\textbf{0.896} & 0.084 & 0.334 & \textbf{<0.01} & \cellcolor{corrHigh}\textbf{0.718} \\
& $MS_{mi}^{ghost}$ & \textbf{<0.01} & \cellcolor{corrHigh}\textbf{0.833} & \textbf{<0.01} & \cellcolor{corrHigh}\textbf{0.747} & \textbf{<0.01} & \cellcolor{corrHigh}\textbf{0.848} & \textbf{<0.01} & \cellcolor{corrVeryHigh}\textbf{0.934} & \textbf{<0.01} & \cellcolor{corrVeryHigh}\textbf{0.963} & \textbf{<0.01} & \cellcolor{corrHigh}\textbf{0.887} & 0.045 & 0.371 & \textbf{<0.01} & \cellcolor{corrModerate}\textbf{0.653} \\
& $MS_{va}^{ghost}$ & \textbf{<0.01} & \cellcolor{corrVeryHigh}\textbf{0.908} & \textbf{<0.01} & \cellcolor{corrVeryHigh}\textbf{0.937} & \textbf{<0.01} & \cellcolor{corrHigh}\textbf{0.810} & \textbf{<0.01} & \cellcolor{corrHigh}\textbf{0.854} & \textbf{<0.01} & \cellcolor{corrVeryHigh}\textbf{0.903} & \textbf{<0.01} & \cellcolor{corrHigh}\textbf{0.787} & 0.021 & 0.410 & \textbf{<0.01} & \cellcolor{corrModerate}\textbf{0.545} \\
& $MS_{ps}^{ghost}$ & 0.118 & 0.311 & 0.052 & 0.363 & \textbf{<0.01} & \cellcolor{corrHigh}\textbf{0.839} & \textbf{<0.01} & \cellcolor{corrVeryHigh}\textbf{0.910} & \textbf{<0.01} & \cellcolor{corrHigh}\textbf{0.900} & \textbf{<0.01} & \cellcolor{corrHigh}\textbf{0.814} & \textbf{<0.01} & \cellcolor{corrModerate}\textbf{0.572} & \textbf{<0.01} & \cellcolor{corrLow}\textbf{0.495} \\
\bottomrule
\end{tabular}}
    \label{tab:muti_corr_test}
    \vspace{-6pt}
\end{table}

\textbf{\textit{Multiple Correlation Analysis}}. 
Table~\ref{tab:muti_corr_test} shows the results of the multiple correlation analysis between mutation scores and the combination of dropout rate and block size. The analysis reveals that the majority of the correlations are statistically significant ($p < 0.01$) and range from high to very high correlation, confirming that the proposed mutation scores are effective metrics capable of capturing uncertainty-induced performance degradation. 
Specifically, the \objms and the \ioums exhibit the strongest correlation with the combined two mutation ratios. For instance, \msiou consistently achieves very high correlation coefficients ($R > 0.90$) across nearly all case studies, such as $R=0.987$ for \scdm{1} and $R=0.977$ for \stdm{5}. Similarly, the \objms maintain high correlation values in most cases, e.g., $0.985$ for \objmsghost in \hfdm{2}. This observation confirms the earlier conclusion that these metrics have clear monotonic gradients and are highly sensitive to the uncertainty in the models. As for \uams metrics, the metrics derived match and miss sets achieve significant and moderate to very high correlations with the mutation ratios, indicating their effectiveness in capturing the uncertainty-induced failures in the models. 
However, the correlation strengths for \uams metrics derived from the ghost set are notably weaker, particularly regarding the \uamsghostps. As the table shows, for \hfdm{1} and \hfdm{2}, the correlation between \uamsghostps and the combined mutation ratios is statistically insignificant ($p > 0.05$), while for \stdm{5}, the correlation is low ($R=0.495$), indicating limited sensitivity of these metrics to uncertainty-induced faults in ghost objects. 
Furthermore, \stdm{4} exhibits insignificant correlations across four of the five ghost-related \uams metrics. These findings are consistent with our previous findings and validate the flat surface phenomenon observed in the descriptive analysis, indicating that the \uams derived from the ghost set are irregular and do not scale linearly with the models' uncertainty.

\begin{center}
    \fcolorbox{black}{gray!10}{\parbox{\textwidth}{\textbf{Conclusion for 
    RQ3}: The proposed mutation scores are generally effective in capturing uncertainty-induced failures of \gls{dl}-enabled software, and are sensitive to varying uncertainty levels. Among all mutation scores, \objms and \ioums metrics are the most effective, showing consistently high to very high correlation with the mutation ratios across both \mcd and \mcb operators, though higher uncertainty can affect their stability. Furthermore, the \uams derived from the match set are the most stable metrics for \mcd operator, while the \uams derived from the match and miss sets are the most stable metrics for \mcb operator.
    }}
\end{center}

\subsection{Data Availability}\label{subsec:data}
To enable the reproducibility of our findings, we provide the replication package in our GitHub online repository\footnote{\url{https://github.com/Simula-COMPLEX/UAMTERS}}. The replication package contains the following contents: 1) the related algorithms and source code for experimenting; 2) the data and code for the experiment results and analysis.

\section{Discussion}

\textbf{\textit{Incorporating Uncertainty in Mutation Analysis}}. In traditional mutation analysis for \gls{dl}-enabled software, mutants are designed to simulate faults in \gls{dl} models; however, they often fail to capture the stochastic behavior inherent in \gls{dl} models. To address this limitation, we propose uncertainty-aware mutation operators that explicitly model uncertainty in \gls{dl} models. Consequently, uncertainty-aware mutation analysis provides a more faithful representation of potential failures that self-adaptive robotic systems may encounter under unpredictable, dynamic conditions, enabling a systematic evaluation of test suite effectiveness in revealing uncertainty-induced model degradation.
Moreover, we introduce uncertainty-aware mutation scores to quantify a test suite's effectiveness in exposing failures induced by stochastic perturbations. Compared to traditional mutation scores, these scores provide a more fine-grained assessment of test suite quality and can further guide the development of tests that enhance safety, reliability, and performance under real-world operating conditions.

\textbf{\textit{Supporting Uncertainty-Aware Mutation Testing}}. While \uamters primarily focuses on mutation analysis of test suites, it can support the mutation testing of \gls{dl}-enabled software. Mutation testing relies on the premise that test suites with high mutation scores are more likely to expose faults in the original program, whereas low mutation scores indicate less effective test suites and therefore highlight the need for improvement. Under this premise, uncertainty-aware mutation testing provides a systematic way to identify high-quality test suites and guide the improvement of low-quality ones that fail to expose uncertainty-induced degradation. Specifically, after evaluating test suite quality via mutation analysis, uncertainty-aware mutation testing can refine or augment low-quality test suites, e.g., by introducing adversarial perturbations~\cite{9438605}, to construct new test suites that more effectively reveal uncertainty-induced degradation. This iterative process enables continuous improvement of test suites that can effectively evaluate the model behaviors under uncertainty. Developing such uncertainty-aware mutation testing techniques is our future work.

\textbf{\textit{Supporting the MAPLE-K Loop for Self-Adaptive Robots}}.
\uamters directly supports the Legitimate component of the MAPLE-K loop for self-adaptive robots. It enables the systematic evaluation of test suites in exposing uncertainty-induced failures, allowing test suites generated by various test generation approaches to be effectively compared. By leveraging these insights, the Legitimate component can make more informed decisions about whether planned adaptations satisfy safety and performance requirements before execution, thereby enhancing the trustworthiness and reliability of self-adaptive robots in real-world operation. 
In addition, the novel mutation scores can further guide and improve test generation. For instance, they can be used to define the objective functions that guide the generation of more effective test suites. Another potential application is to support test prioritization, ensuring that higher quality tests are executed first to efficiently detect potential failures.
Beyond the Legitimate component, \uamters can also support other components of the MAPLE-K loop. For example, the Knowledge component can store model uncertainty and test suite quality information, enriching the system's historical knowledge for future decision-making. Similarly, insights from \uamters could inform the Analyze component by highlighting uncertainty-related failure patterns, and guide the Plan component to generate more robust adaptation strategies.

\section{Threats to Validity}
\textit{Conclusion Validity} relates to the validity and reliability of the conclusions. To draw reliable conclusions, we first adopt the statistical definition of mutation killing proposed by Jahangirova and Tonella~\cite{jahangirova2020empirical}, which accounts for the stochastic nature of DL models. In addition, we employ appropriate statistical tests and conduct rigorous analysis in accordance with established guidelines.

\textit{Construct Validity} concerns the definition of mutation scores. The scores are defined based on the well-established performance metrics and uncertainty metrics, e.g., IoU and Entropy, commonly used in evaluating DL-based systems. This ensures that the mutation scores reflect test suite effectiveness in revealing performance degradation under uncertainty.

\textit{Internal Validity} concerns the parameter settings. To determine the mutation ratios for the mutation operators, we follow the dropout rate settings from Lu et al.~\cite{lu2025assessing} and the block size settings from Yelleni et al.~\cite{yelleni2024monte} to select the mutation ratios. Besides, the selection of network layers to which the Dropout or DropBlock layers are applied may influence the observed mutation effects. To reduce this threat, we follow common practice in uncertainty quantification~\cite{catak2021prediction,lu2025assessing} by applying the mutation operators to the last three convolutional layers.

\textit{External Validity} is about the generalizability of this study. First, the selection of the subject robotic systems can potentially threaten the external validity. To mitigate this threat, we select three industrial-level robotic systems comprising seven \gls{dl}-enabled object detection software and eight test suites representative of real-world, safety-critical applications.
Second, we admit that the mutation scores are designed for object detection tasks, which may limit their direct applicability to other DL tasks or model architectures. Nevertheless, focusing on object detection remains essential, as it's a core perception task in robots, supporting safety-critical functions such as navigation, obstacle avoidance, and human–robot interaction. Dedicated mutation scores, therefore, enable a more accurate evaluation of test suite effectiveness in revealing uncertainty-induced degradation. Besides, while these metrics are designed for object detection models, the underlying uncertainty-aware mutation testing methodology is general and can be adapted to other DL tasks.

\section{Related Work}

Mutation analysis is a widely applied technique for evaluating test suite quality for software systems and can support a range of activities, including test generation~\cite{haga2012automatic,baker2012empirical}, test prioritization~\cite{shin2019empirical}, and fault localization~\cite{papadakis2014effective,moon2014ask}. In recent years, mutation analysis has been extended to \gls{dl}, with specialized mutation operators developed to simulate realistic faults in neural networks. These operators can target various aspects of \gls{dl} models, including weights, neuron activations, layer connections, and input data~\cite{shen2018munn,ma2018deepmutation,zhang2020machine}. For example, MuNN~\cite{shen2018munn} is a mutation analysis tool which integrates five mutation operators covering both domain-dependent and depth-dependent mutations for \gls{dl}.
DeepMutation~\cite{ma2018deepmutation} proposes to inject faults into \gls{dl} models using both source-level and model-level mutation operators. Later, DeepMutation++~\cite{hu2019deepmutation++} extended DeepMutation by introducing mutation operators specifically tailored for recurrent neural networks. 
Jahangirova and Tonella~\cite{jahangirova2020empirical} conduct an empirical evaluation to identify effective mutation operators and introduce a statistical definition of mutation killing. To simulate real DL faults, Humbatova et al.~\cite{humbatova2021deepcrime} propose DeepCrime, which integrates 35 mutation operators by analyzing existing real DL faults. Although DeepCrime includes a mutation operator that modifies dropout layers, it is designed as a regularization operator applied during training and does not explicitly model predictive uncertainty or systematically assess uncertainty-induced behavioral changes. In contrast, our dropout-based operator is designed as an uncertainty-aware mutation operator applied at prediction time, enabling the explicit modeling and evaluation of predictive uncertainty.

Moreover, mutation analysis has been applied to guide mutation testing. For example, Tambon et al.~\cite{tambon2023probabilistic} introduce PMT, a probabilistic mutation testing approach that accounts for stochastic model behavior to provide more consistent and reliable mutation-based test evaluations in DL models. Wang et al.~\cite{wang2021prioritizing} employ mutation analysis to prioritize test inputs for DL models, improving the effectiveness of test suites in revealing model weaknesses. Dang et al.~\cite{dang2023graphprior} propose GraphPrior, a mutation-based approach for graph neural networks that ranks test inputs by the number of mutated models they kill, improving fault detection on both natural and adversarial inputs. Wang et al.~\cite{8812047} propose a runtime detection method for adversarial examples based on mutation testing, demonstrating that adversarial inputs exhibit higher sensitivity to model mutations than normal inputs. Pour et al.~\cite{9438605} propose a search-based testing framework for \gls{dl} in source code embedding tasks, where mutation testing is used to guide adversarial test generation.

While these works have advanced mutation analysis and testing for \gls{dl}, they do not explicitly address failures related to model uncertainty, which is crucial in safety-critical applications such as robotics~\cite{lu2025assessing} and autonomous driving~\cite{michelmore2020uncertainty}. In this paper, we specifically focus on model failures that are related to model uncertainty and propose two uncertainty-aware mutation operators to inject uncertainty into the models. In addition, we design 19 novel mutation scores that capture different perspectives of model behavior, enabling a more comprehensive assessment of test suite effectiveness in detecting uncertainty-related performance degradation.

\section{Conclusion}
To evaluate test suite quality, we present \uamters, an uncertainty-aware mutation analysis framework for DL-enabled software in robots. \uamters introduces novel mutation operators that explicitly inject stochastic uncertainty into DL models, enabling the systematic simulation of uncertainty-induced behaviors. In addition, we define novel uncertainty-aware mutation scores to quantify a test suite's ability to reveal model degradation and behavioral changes under varying levels of uncertainty. The empirical evaluation on three industrial-level robotic case studies demonstrates that \uamters more effectively distinguishes test suite quality and captures uncertainty-induced failures compared to traditional mutation scores. By explicitly considering uncertainty, \uamters provides a more informative evaluation of test suite effectiveness, supporting the validation of dependability in self-adaptive robots. In the future, we plan to extend \uamters by integrating additional uncertainty-aware mutation operators and metrics designed for more complex \gls{dl} models, including vision language action models. We also aim to investigate strategies for integrating \uamters into continuous testing pipelines for self-adaptive robots, enabling real-time assessment of test suite quality under evolving operational conditions. Furthermore, studying the impact of different types of uncertainty on test effectiveness could provide deeper insights into designing robust and resilient robots.

\section*{Acknowledgments}
% \begin{acks}
This work is supported by the RoboSAPIENS project funded by the European Commission's Horizon Europe programme (No. 101133807). Jiahui Wu is also supported by the Co-tester (No. 314544) project funded by the Research Council of Norway.
% \end{acks}

%Bibliography
\bibliographystyle{unsrt}  
\bibliography{references}

\end{document}